\documentclass[aps,pra,twocolumn,epsfig,floats,superscriptaddress]{revtex4}
\usepackage{amsmath}
\usepackage{color}
\usepackage{graphicx}
\usepackage{amsfonts}
\usepackage{mathrsfs}
\usepackage{float}
\usepackage{amsmath}
\usepackage{amssymb}
\usepackage{physics}
\usepackage[utf8x]{inputenc}
\usepackage{times}
\usepackage{hyperref}
\hypersetup{
  colorlinks=true,
  citecolor=blue,
  linkcolor=blue,
  urlcolor=blue}

\DeclareSymbolFont{rmlargesymbols}{OMX}{mdbch}{m}{n}
\DeclareMathSymbol{\rmintop}{\mathop}{rmlargesymbols}{82}

\begin{document}

\title{Formation Dynamics of Quantum Droplets for Homonuclear and Heteronuclear Mixtures}

\author{Enrique Calderoli}
\email{enrique.calderoli@gmail.com}
\affiliation{Instituto de F\'{\i}sica, Universidade Federal do Rio Grande do Sul, Porto Alegre-RS, 91501-970, Brasil}
\affiliation{Departamento de Psiquiatria, Universidade Federal do Rio Grande do Sul, Porto Alegre-RS, 90035-903, Brasil}

\author{Gerardo Mart\'{\i}nez}
\email{martinez@if.ufrgs.br} 
\affiliation{Instituto de F\'{\i}sica, Universidade Federal do Rio Grande do Sul, Porto Alegre-RS, 91501-970, Brasil}

\date{\today}

\begin{abstract}
Significant efforts have been devoted to studying the properties of quantum droplets, an ultra low-temperature phase of bosonic quantum matter that emerges as a consequence of the Lee-Huang-Yang fluctuating correction. However, the temporal dynamics of droplet formation for heteronuclear bosonic mixtures is only partially understood. Here, we numerically analyze the droplet formation process for homonuclear and heteronuclear binary bosonic mixtures in one dimension, using a tight-binding model and real-time evolution with a novel, highly robust integration algorithm. We proceed with a systematic scan of interaction intensities, mass ratios, and initial conditions that allows us to characterize quantitative criteria for droplet formation and equilibrium prospects. Noticeably, most droplets readily form across the entire parameter space, although only a small fraction achieves a stable equilibrium configuration within the simulation horizon. We attribute this equilibrium deficiency to damping from a breathing mode, which we extract directly from the width oscillations at late times. The Lee-Huang-Yang contribution supplies essentially the entire binding energy, while, at late times, the density profile of the equilibrated subset is better described by a soliton-like shape rather than the flat-topped profiles characteristic of larger droplets. Referring to the energy of the free-atom band, the binding energy grows super-extensively with the number of atoms as $E_{\text{bind}} \propto N^{1.6}$. Heteronuclear droplets bind more strongly as the mass ratio between their components increases and exhibit breathing oscillations that are greater than those of their homonuclear counterparts, which is consistent with the role of mass-imbalanced kinetic terms. Our analysis provides a methodological framework for interpreting real-time quantum droplet simulations.
\end{abstract}

\maketitle
\section{Introduction}
\label{introduction}
Petrov's prediction \cite{petrov2015prediction} of a new ultracold phase of bosonic quantum matter stabilized by the Lee-Huang-Yang (LHY) fluctuating corrections to the mean-field  model triggered a huge torrent of investigations, both theoretical and experimental, into the nature and dynamics of this self-bound state, termed a quantum droplet due to its hydrodynamic properties. In the last decade, these quantum droplets have been observed in dipolar gases of dysprosium and erbium \cite{kadau2016observing,ferrier2016observation,schmitt2016self,chomaz2016quantum}, as well as in binary mixtures of potassium isotopes \cite{cabrera2018quantum,semeghini2018self,cheiney2018bright}.

The established analytical consensus states that quantum droplet formation occurs as a result of scaling differences between mean-field terms and LHY corrections to the energy density expression. In 3D bosonic systems where droplets can form, the mean-field terms have different signs (intraspecies being repulsive and interspecies being attractive), largely canceling each other out, with a net effect scaling with $-n^{2}$, where $n$ is the local atomic density. The possibility of droplet states arises because the repulsive LHY correction for this system scales with $n^{5/2}$ \cite{petrov2015prediction}, compensating for the net result of the mean-field terms,  creating a local minimum in the system's energy curves, and preventing the usual collapse. In contrast, 1D Bose gases led to droplet formation with an inverse configuration: a liquid mean-field effect being repulsive and scaling with $n^{2}$ and a collapse being prevented due to attractive LHY corrections scaling with $-n^{3/2}$ \cite{petrov2016ultradilute,parisi2019liquid,bottcher2021new,luo2021new}. The particularities of this dimensional crossover have been the subject of many recent and ongoing research efforts \cite{edler2017quantum,tylutki2020collective,edmonds2020quantum}.

Despite the stream of studies on droplet phenomena, many points about this phase, in particular its dynamical properties, have not yet been adequately addressed \cite{cavicchioli2025dynamical,richaud2019mixing,richaud2019pathway}. For example, the theoretical formulation of droplets generally assumes homonuclear 
components, refraining from modeling heteronuclear systems due to analytical and computational challenges in dealing with different masses in the Bogoliubov treatment of quantum excitations \cite{mistakidis2021formation,d2019observation}. Furthermore, most studies typically employ an approximation for the LHY correction, whose form depends on the specific geometry being studied and is usually taken as some power of $n$, the local atom density. Clearly, this leaves a large fraction of bosonic systems unaccounted for when it comes to physical models. Another gap in our understanding of droplet physics stems from the fact that most investigations employ imaginary-time evolution techniques, thus analyzing these liquid-like states only when equilibrium is reached. Hence, a wealth of physical information about the  dynamics of quantum droplets remains unexplored.

In this work, we study the real-time formation dynamics of quantum droplets in two-component entangled Bose-Einstein condensates (BECs) on a 1D discrete lattice using a tight-binding formulation. We analyze how the consideration of different masses for each component 
alters the systems' dynamics from the homonuclear picture and explores the impact of initial preparation states on the formation and stabilization of the droplets. 

The remainder of the paper is structured as follows. In Sect. \ref{theoretical_framework}, we present the model used and the variational method employed based on a Bogoliubov formulation that includes first-order LHY energy corrections. The main point addressed here is how to handle numerically the chemical potential resulting from the LHY energy part. Some details of the numerical scheme implementation are left to the Appendices. In Sect. \ref{section-III}, after a brief review of the Gross-Pitaevskii equations used in this work,  we present the most relevant findings on the dynamical formation of droplets, contrasting the differences between homonuclear and heteronuclear cases. To this end, we analyze the time formation and equilibration of droplets, the energy partitioning and binding dynamics, the localization properties, the component overlap and coalescence dynamics, the oscillating breathing modes, and the density profiles. Finally, Sect. \ref{conclusions} sets out the conclusions that we can arrive at from the data and analysis of this study.

\section{Theoretical Framework}
\label{theoretical_framework}
The system we study is a two-component Bose mixture in 1D whose Hamiltonian in real-space representation is given by
its second quantized form as
\begin{multline}
    \hat H = \int dx \left[ \,
\sum_{\sigma=1,2} \hat\Psi_\sigma^\dagger(x)\left(-\frac{\hbar^2}{2m_\sigma}\frac{d^2}{dx^2}\right)\hat\Psi_\sigma(x) \right. \\ \\
\left. + \sum_{\sigma=1,2}\frac{U_\sigma}{2}\hat\Psi_\sigma^\dagger \hat\Psi_\sigma^\dagger \hat\Psi_\sigma \hat\Psi_\sigma
+ U_{12}\,\hat\Psi_1^\dagger\hat\Psi_2^\dagger\hat\Psi_1\hat\Psi_2
\right],
\end{multline}
where $\hat\Psi_\sigma(x)\,(\hat\Psi^\dagger_\sigma(x))$ annihilates (creates) a boson of species $\sigma$ at position $x$, $m_1 (m_2)$ is the atomic mass for the first (second) condensate.\ $U_1 (U_2)$ is the intraspecies atomic interaction for the first (second) condensate and $U_{12}$ is the interspecies atomic interaction, all assumed to be short range \cite{pitaevskii2016bose,pethick2008bose}.

This Hamiltonian is a useful model for ``quasi-one-dimensional” binary bosonic mixtures, referred to as such because, although they are physically three-dimensional in the laboratory, their dynamics are effectively one-dimensional due to a strong confinement \cite{olshanii1998atomic,dunjko2001bosons} in the transverse directions. This can be achieved through the application of a very tight harmonic trap in two spatial directions, while leaving the third relatively free. When the transverse trapping frequency $\omega_{\bot}$ is sufficiently large so that the associated energy scale $\hbar\omega_{\bot}$ exceeds the other relevant energies of the system, such as the chemical potential $\mu$, the thermal energy $k_{B}T$, and the interaction energy, the atoms are frozen in the transverse ground state \cite{gorlitz2001realization}, unable to access the excited transverse modes. As a result, all its relevant dynamics are confined to the axial direction. This condition will be assumed in our study.

The distinction between the condensate phase and the excited states is most easily treated within the context of momentum-space representation, which we adopt from now on. We begin by expressing the field operators on a plane-wave basis in momentum space in the following form:
\begin{align}
\hat\Psi_1(x) &= \frac{1}{\sqrt{L}}\left(\hat a_0 + \sum_{k\neq 0} e^{ikx}\hat a_k\right),\\
\hat\Psi_2(x) &= \frac{1}{\sqrt{L}}\left(\hat b_0 + \sum_{k\neq 0} e^{ikx}\hat b_k\right),
\end{align}
where $k=2\pi n/L$ with $n\in\mathbb{Z}$ and $\hat a_k$ ($\hat b_k$) annihilates a boson of atom species $1$ ($2$) with momentum $\hbar k$ \cite{bogoliubov1947theory}. As such, in both equations above, the first term in parentheses represents the bosons composing the condensate, and the second term represents the bosons at excited states.

The Bogoliubov approximation is based on the assumption that the condensate mode is macroscopically occupied, making $\hat a_0$ and $\hat b_0$ massive compared to the small fluctuations represented by $\hat a_{k\neq 0}$ and $\hat b_{k\neq 0}$ \cite{bogoliubov1947theory,fetter2012quantum}. Thus, when expanding the expression for the Hamiltonian in momentum representation, one keeps only terms up to quadratic order, which results in:
\begin{equation}
    \hat H = \hat H^{(0)} + \hat H^{(2)} + \mathcal{O}(\hat a_k^3,\hat b_k^3),
\end{equation}
where $\hat H^{(0)}$ contains only condensate operators and
$\hat H^{(2)}$ is bilinear in the nonzero-$k$ operators and encodes the elementary excitations.

The standard mean-field (MF) treatment consists of neglecting the excited states altogether and equating the condensate mode operators to $c$-numbers ($\hat a_0 \to \sqrt{N_A},\; \hat b_0 \to \sqrt{N_B}$), which leads to the MF energy density:
\begin{equation}
    \varepsilon_{\rm MF}=\frac{E_{\rm MF}}{L}
= \frac{1}{2}U_1 n_1^2 + \frac{1}{2}U_2 n_2^2 + U_{12} n_1 n_2,
\end{equation}
where $n_\sigma=N_\sigma/L$ \cite{dalfovo1999theory}. 

Considering that quantum fluctuations only slightly deplete the condensate, the first-order correction to the mean-field treatment is reached when one accounts for the condensate depletion by setting
\begin{equation}
    N_1 = \hat a_0^\dagger \hat a_0 + \sum_{k\neq 0}\hat a_k^\dagger \hat a_k, \,\,\,\, N_2 = \hat b_0^\dagger \hat b_0 + \sum_{k\neq 0}\hat b_k^\dagger \hat b_k,\,
\end{equation}
which can be inverted to express the ground-state operators as functions of the operators $N_1$ and $N_2$ and of the excited states. After this step, the Hamiltonian becomes a sum over independent $(k,-k)$ sectors.

Defining the Nambu spinor as $\Phi^\dagger = \big(\hat a_k^\dagger,\ \hat b_k^\dagger,\ \hat a_{-k},\ \hat b_{-k}\big)$ \cite{de2018superconductivity}, the quadratic Hamiltonian can be compactly written as
\begin{multline}\label{hamiltonian}
\hat H
= \sum_{k>0} \Phi^\dagger\, \mathcal{H}\,\Phi
+ \frac{U_1 N_1^2}{2L}+\frac{U_2 N_2^2}{2L}+\frac{U_{12}N_1 N_2}{L}\\\\
-\sum_{k>0}\left(\frac{\hbar^2 k^2}{2m_1}+\frac{\hbar^2 k^2}{2m_2}+U_1 n_1+U_2 n_2\right).
\end{multline}
Here, the sum is over $k>0$ only, as each positive $k$ represents the pair $(k,-k)$. The $4\times 4$ matrix $\mathcal{H}$ in Eq. (\ref{hamiltonian}) is
\begin{equation}
\mathcal{H} =
\begin{pmatrix}
h_1(k) & h_{12} & U_1 n_1 & h_{12}\\
h_{12} & h_2(k) & h_{12} & U_2 n_2\\
U_1 n_1 & h_{12} & h_1(k) & h_{12}\\
h_{12} & U_2 n_2 & h_{12} & h_2(k)
\end{pmatrix},
\end{equation}
where
\begin{equation}
h_\sigma(k) = \frac{\hbar^2 k^2}{2m_\sigma}+U_\sigma n_\sigma,
\quad
h_{12}=U_{12}\sqrt{n_1 n_2}.
\end{equation}
Because $\Phi$ contains both creation and annihilation operators, diagonalizing the quadratic Hamiltonian is not a standard unitary diagonalization, instead one must solve a Bogoliubov-de Gennes (BdG) type eigenproblem. Formally, we introduce the commutation (metric) matrix
\begin{equation}
  M_b = [\Phi,\Phi^\dagger] = \begin{pmatrix} I_2 & 0\\ 0 & -I_2\end{pmatrix},  
\end{equation}
and the excitation frequencies $\omega$ are obtained \cite{blaizot1988quantum} by solving the characteristic equation
\begin{equation}
\det\!\left(M_b H - \omega\, I_4\right)=0.
\end{equation}
In this two-component case, the characteristic polynomial reduces to a quadratic equation for $\omega^2$, which yields two excitation branches:
\begin{equation}
    \omega_{\pm}^2 = \frac{\varepsilon_1^2 + \varepsilon_2^2}{2} \pm \sqrt{\frac{(\varepsilon_1^2 - \varepsilon_2^2)^2}{4} + U_{12}^2n_1n_2\frac{\hbar^4k^4}{m_1m_2}}\,,
\end{equation} 
where
\begin{equation}
    \varepsilon_{\sigma} = \left[\frac{\hbar^2k^2}{2m_{\sigma}}\left(\frac{\hbar^2k^2}{2m_{\sigma}} + 2U_{\sigma}n_{\sigma}\right)\right]^{1/2}. 
\end{equation}
Through a Bogoliubov transformation, the Hamiltonian in Eq. \eqref{hamiltonian} can be rendered into the ground state energy density of the system with the first-order LHY correction included \cite{lee1957eigenvalues}:
\begin{multline}\label{energy-density-ground-state}
    \varepsilon = \frac{E_0}{L} = \frac{U_1n_1^2}{2} + \frac{U_2n_2^2}{2} + U_{12}n_1n_2 \\\\ +\frac{1}{L}\sum_{k>0}\left(\omega_{+} + \omega_{-} - \sum_{\sigma=1,2} \left( \frac{\hbar^2k^2}{2m_\sigma} + U_\sigma n_\sigma \right)\right).
\end{multline}
Employing a Euler-Lagrange framework to this energy density expression leads to the extended Gross-Pitaevskii equations \cite{gross1961structure,pitaevskii1961vortex} for the system:
\begin{equation}
    i\hbar\frac{\partial \Psi_{\sigma}}{\partial t} = \left(-\frac{\hbar^2}{2m_{\sigma}}\frac{\partial^2}{\partial x^2} + U_{\sigma}n_{\sigma} + U_{12}n_{\sigma'\neq\sigma} + \Delta\mu_{\sigma}^{\text{LHY}}\right)\Psi_{\sigma},
\end{equation}
where the chemical potential terms due to the LHY correction are obtained from 
\begin{equation}
    \Delta\mu_{\sigma}^{\text{LHY}} = \frac{\partial\varepsilon^{\text{LHY}}}{\partial n_{\sigma}},
\end{equation}
with $\sigma=1,2$, where $\varepsilon^{\text{LHY}}$ is the contribution of the LHY correction to the energy density of the ground state of the system as a whole, and is given by the terms inside the summation over the states $k>0$ in Eq. \eqref{energy-density-ground-state}. 

To evaluate this sum for this one-dimensional system, we can replace such a discrete sum by an integral in momentum-space, which gives the following
\begin{equation}
    \varepsilon^{\text{LHY}} =\int_0^{+\infty} \frac{dk}{2\pi}\left(\omega_{+} + \omega_{-} -\! \sum_{\sigma=1,2}\left(\frac{\hbar^2k^2}{2m_\sigma} + U_\sigma n_\sigma\right)\right).
    \label{energy-density-integral}
\end{equation}
Considering the variational approach used above, we find that the chemical potential contribution due to the LHY correction is given by
\begin{equation}
\label{chemical-potential-integral-in-k}
    \Delta\mu_{\sigma}^{\text{LHY}} = \int_0^{+\infty}\frac{dk}{2\pi}\left(\frac{\partial\omega_{+}}{\partial n_{\sigma}} + \frac{\partial \omega_{-}}{\partial n_{\sigma}} - U_{\sigma}\right).
\end{equation}
So far, most studies on quantum droplets were restricted to studying homonuclear ($m_{1} = m_{2} \equiv m$) Bose gases mixtures under the single-mode approximation ($n_{1} = n_{2} \equiv n$) near the critical threshold $\vert U_{12}\vert\approx \sqrt{U_{1}U_{2}}$ \cite{petrov2015prediction,astrakharchik2018dynamics}, which implies a residual value $\delta U \equiv U_{12} + \sqrt{U_{1}U_{2}} \approx 0$. In particular, $\delta U< 0$ means $ U_{12} < -\sqrt{U_1U_2}$ in that case. In the critical region with $\delta U \approx 0$, where the LHY correction becomes relevant, droplet formation has been widely demonstrated, so we shall look for the effect of the values of $\delta U$ with a wider scope.

\section{Homonuclear versus heteronuclear systems}
\label{section-III}
In this Section, we analyze how employing different masses for the species forming condensates 1 and 2 impacts the state and dynamics of the quantum droplets. We also investigate the effect of different initial configurations on the system's behavior and constitution. We do so by numerically integrating the following set of coupled extended Gross-Pitaevskii equations:
\begin{align}
i\hbar \frac{\partial \Psi}{\partial t} &= -\frac{\hbar^2}{2m_1}\frac{\partial^2\Psi}{\partial x^2} + \Bigl(U_1|\Psi|^2 + U_{12}|\Phi|^2 
 + \Delta\mu_1^{\text{LHY}}\Bigr)\Psi,  \\[1ex]
i\hbar \frac{\partial \Phi}{\partial t} &= -\frac{\hbar^2}{2m_2}\frac{\partial^2 \Phi}{\partial x^2} + \Bigl(U_2|\Phi|^2 + U_{12}|\Psi|^2  + \Delta\mu_2^{\text{LHY}}\Bigr)\Phi, 
\end{align}
for a binary Bose mixture in 1D. Here, $\Psi(x,t)$ represents the wavefunction of the first condensate, with bosons of mass $m_1$, while $\Phi(x,t)$ represents the wavefunction of the second condensate, composed of bosonic particles of mass $m_2$. We study how this system evolves for a tight-binding model with nearest-neighbor hopping on a discrete lattice of 513 sites, using a variation of the Crank-Nicolson method to solve this coupled set of equations. The units for the results shown below are given in terms of the hopping of the first condensate and use $\hbar=1$. The integral of Eq. \eqref{chemical-potential-integral-in-k} for the chemical potential contributions of the LHY correction to the first  and second condensate is calculated using the exp-sinh quadrature, a variant of the tanh-sinh quadrature. As is common in the literature \cite{englezos2025multicomponent}, we neglect the dynamical instabilities that occur from imaginary values for chemical potentials. Details of the numerical implementation can be found in the Appendices \ref{Crank-Nicolson appendix} and \ref{tanh-sinh-quadrature}.

 We consider four types of initial conditions across two modalities: (i) initial shape, with condensates starting out in either a Gaussian configuration or a delta-like configuration, and (ii) initial separation, with condensates departing from the same site on the lattice (no separation), or from distinct sites (variable separation). We study how the variation of these conditions affects the dynamics of the droplets.

\subsection{Formation Time and Equilibrium Time}\label{sec:formation_time}
To classify whether or not a simulation formed a droplet-like structure, we define two criteria: (i) a localized density profile at the end of the simulation, quantified as a ratio of peak density to average density at the edges of the lattice that exceeds a threshold of 10 times; and (ii) a negative chemical potential in the core region, which is defined as the set of points at which $n > 0.7n_{\text{peak}}$. The formation time $t_{\text{form}}$ is therefore the first occurrence of the simulation in which both criteria are met. The measurement window for the chemical potential criterion was the final 20\% of the simulation. We deliberately do not use the total energy as a formation criterion: in a network the total energy is conserved and bounded below by the free-atom band energy, so its sign is fixed by preparation of the initial state rather than by the dynamical outcome; the binding energy is reported instead as a physical property of each run as deployed in Section \ref{sec:energy-partition-binding}.
 
A second, more stringent test is equilibration. As we study the formation of droplets and neglect dynamical instabilities that may arise from imaginary terms in the chemical potential contributions from the LHY correction, it is to be expected that few of the simulations will actually reach a very steady state of equilibrium. Nevertheless, we define the formed structure to be equilibrated if, in the final 30\% of the simulation, it satisfies two more conditions: (i) the oscillation amplitude does not exceed 30\% of the mean width, and (ii) the mean width fluctuates by less than 2\%. 
 
\begin{center}
\begin{figure}
\includegraphics[width=\columnwidth]{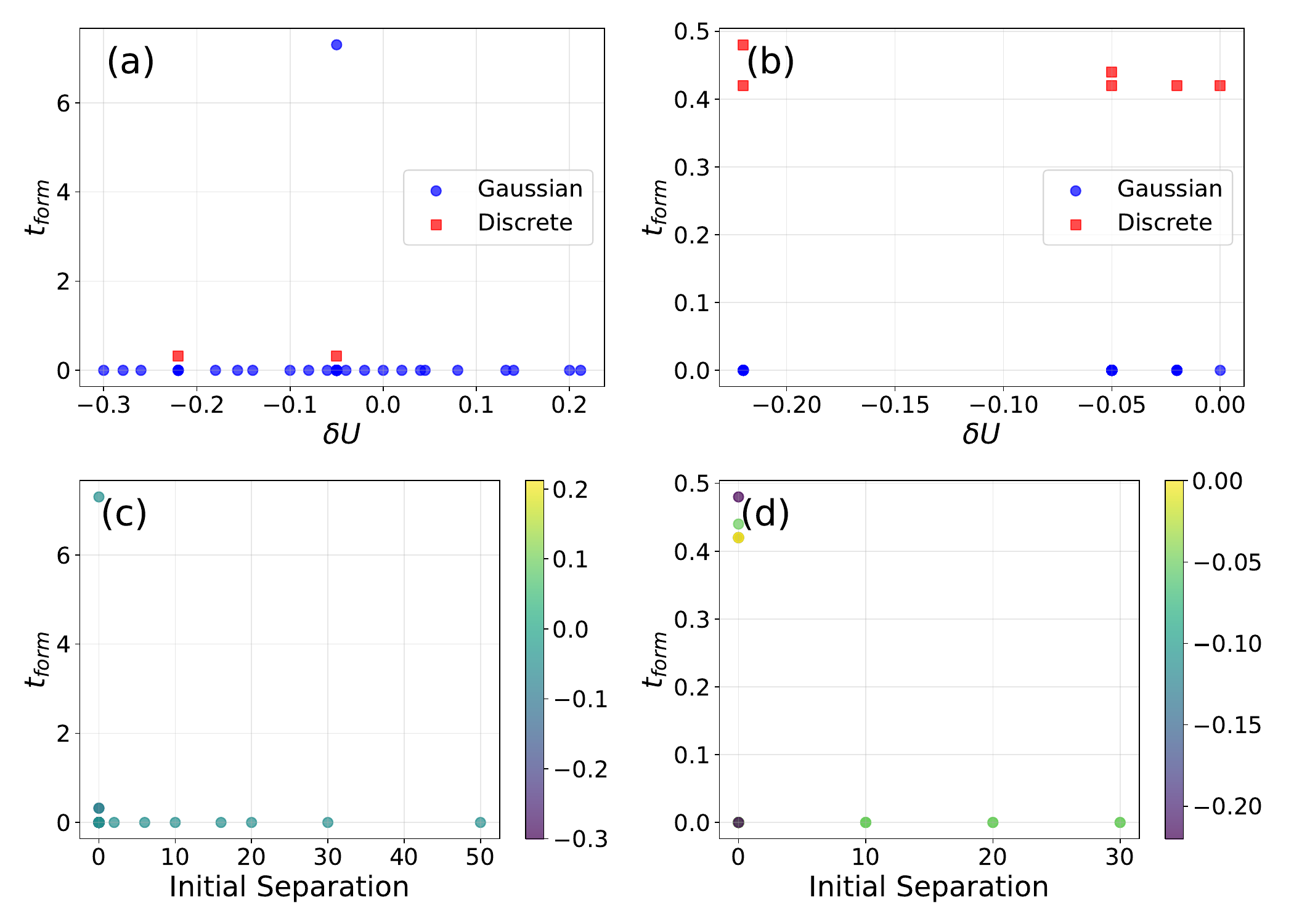}
\caption{Droplet formation time $t_{\text{form}}$ versus $\delta U$ for (a) homonuclear and (b) heteronuclear cases;  $t_{\text{form}}$ versus initial separation for (c) homonuclear and (d) heteronuclear cases. Colored ruling bars are values of $\delta U$. Gaussian initial conditions (blue dots) achieve essentially instantaneous formation of droplets, while discrete initial conditions (red squares) require $t_{\text{form}} \approx 0.4$, in arbitrary units.}
\label{fig:formation_time}
\end{figure}
\end{center}

Figure \ref{fig:formation_time} presents the formation time analysis for homonuclear and heteronuclear systems, as a function of interaction strength variations $\delta U$, as well as initial separation.

Of the 63 homonuclear simulations, 54 resulted in a formed droplet, a success rate of 85.7\%. This is comparable to the 88.6\% success rate for heteronuclear droplet formation (39/44), indicating that mass asymmetry is not a barrier to droplet formation.

Formation is essentially instantaneous for Gaussian initial conditions in both nuclear types ($t_{\text{form}}^{\text{(homo)}} = 0.14 \pm 1.0$ and $t_{\text{form}}^{\text{(hetero)}} = 0.0 \pm0.0$), with all but one of the homonuclear Gaussian cases already satisfying the formation criteria at $t = 0$, indicating that a system with a smooth Gaussian profile merely evolves to confirm the sustained nature of the self-bound state, regardless of mass ratio. Discrete initial conditions require $t_{\text{form}}^{\text{(homo)}} = 0.32 \pm 0.0$ and $t_{\text{form}}^{\text{(hetero)}} = 0.433\pm0.024$ for the localized density to be redistributed into a smooth droplet profile. The formation time does not show dependence on $\delta U$ for the initial Gaussian conditions (Figure \ref{fig:formation_time}, top row), suggesting that the formation criteria capture the preparation of the initial state rather than dynamical condensation.

\vspace{-20pt}
 
When we consider the impact of initial conditions on the success rate of droplet formation, we further establish the importance of the chosen spatial configuration: a total of 107 simulations were attempted, leading to 93 droplets being formed, according to the criteria established above. However, all 85 of the initial Gaussian conditions (100\%) lead to successful droplet formation, while only $36.4\%$ (8/22) of the discrete initial conditions met the formation criteria. This indicates that delta-like initial configurations are somewhat unphysical for modeling structures bound by the LHY correction. 
 
For runs with initially separated components, formation time remains near zero even for initially separated configurations by up to 50 lattice sites, as demonstrated on the bottom row of Figure \ref{fig:formation_time}. The coalescence dynamics (Section \ref{sec:overlap}) proceed independently of the formal formation time.
 
Only 15 of the 93 droplets that were successfully formed reached the equilibrium state at the end of the simulation ($t_{\text{max}} = 125$), as defined by the stated criteria. These criteria are deliberately demanding: a droplet can be well-formed and persist indefinitely, yet fail to meet equilibration thresholds if it retains significant collective excitations. The $83.9\%$ of droplets formed that do not equilibrate need not be unstable, as simply maintaining breathing oscillations or slow drifts is sufficient to prevent them from satisfying the stability criteria.
 
\begin{figure}[ht]
\centering
\includegraphics[width=\columnwidth]{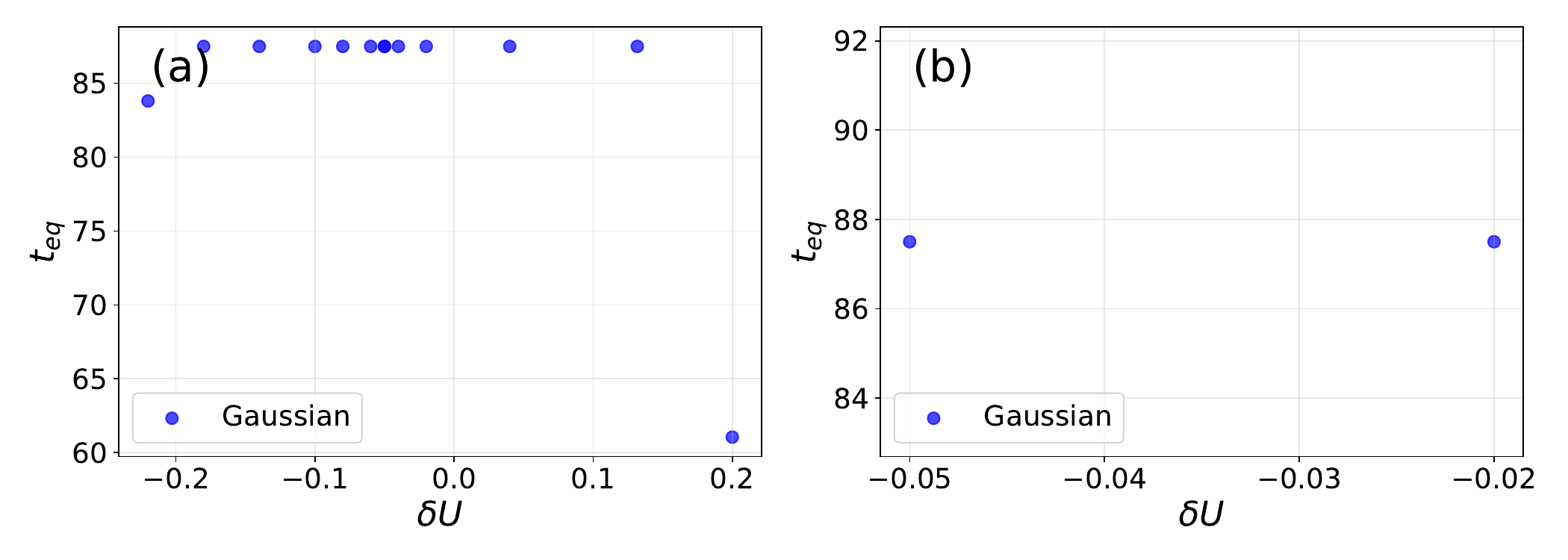}
\caption{Equilibration time $t_{\text{eq}}$ versus $\delta U$ for (a) homonuclear and (b) heteronuclear droplets. Notice a clustering at $t_{eq} \approx 87.5$. See text.}
\label{fig:equilibrium_time}
\end{figure}
 
If a particular simulation has equilibrated at the end of the simulated timescale, we can define the equilibrium time $t_{\text{eq}}$ as the last moment of the simulation in which the width variation exceeded 2\%. Figure \ref{fig:equilibrium_time} presents the equilibration time for homonuclear and heteronuclear droplets as a function of the interaction strength $\delta U$.
Only 24.1\% of the homonuclear droplets formed (13/54) achieve equilibrium within the simulated time scale. This rate is much higher than the success rate for the equilibration of the heteronuclear droplets formed: 5.1\% (2/39), indicating that mass asymmetry represents a crucial factor in relaxation dynamics, preventing or significantly slowing it. The low equilibration rate reflects the persistence of breathing oscillations in 1D systems. Many runs exhibit sustained width variations that, while not growing unboundedly, fail to damp below the 30\% amplitude threshold. This is consistent with the theoretical expectations of undamped collective modes in reduced-dimensional systems \cite{tylutki2020collective}.

The accumulation of equilibration times around $t = 87.5$ in Figure \ref{fig:equilibrium_time} is a statistical artifact of the equilibration threshold used: the final 30\% of the simulation is devoted to the equilibration analysis, so the equilibration time is searched backward from the time point marking 70\% of the timescale, which is $t = 87.5$.

All 15 equilibrated simulations have Gaussian initial conditions and no initial separation. Although the equilibrium criteria place conditions on the shape dynamics of the formed structures, they place no conditions on its energy. As shown below, the total energy is conserved to high accuracy in every run, so there is no independent energetic relaxation channel for such criteria to probe.
 
\subsection{Energy Partitioning and Binding Dynamics}\label{sec:energy-partition-binding}

Understanding the energy budget of quantum droplets is essential for identifying the binding mechanism. The total energy of a droplet is the sum of its kinetic energy, due to the hopping terms of both condensates, its self-energy, due to the intraspecies interactions of both condensates, its mean-field interaction energy, due to the interspecies interaction, and its LHY energy:
\begin{equation}
E_{\text{total}} = E_{\text{hop}}^{(1)} + E_{\text{hop}}^{(2)} + E_{\text{intra}}^{(1)} + E_{\text{intra}}^{(2)} + E_{\text{inter}} + E_{\text{LHY}}
\label{energy-components}
\end{equation}
This expression provides a fundamental characterization of the overall stability of the droplet.
 
On a lattice, the natural zero of energy is not zero but the band energy of free atoms: a single delocalized particle of species $\sigma$ has minimum energy $-2J_\sigma$, minus twice its hopping energy, so $N_1+N_2$ fully dissociated atoms at rest carry
\begin{equation}
E_{\text{free}} = -2\,(J_1 N_1 + J_2 N_2),
\qquad
E_{\text{bind}} = E_{\text{total}} - E_{\text{free}},
\label{eq:ebind}
\end{equation}
and a structure is energetically bound if and only if $E_{\text{bind}} < 0$. All binding energies quoted below refer to Eq.~\eqref{eq:ebind}. A second structural fact organizes this section: the total energy is a constant of motion. Across all 107 runs, the difference between the late-time average and the global minimum of $E_{\text{total}}$ has a median of $1.4\times10^{-3}$ and never exceeds $5.2\times10^{-2}$, which limits the drift of the semi-implicit time integration. Consequently, $E_{\text{total}}$ and $E_{\text{bind}}$ are fixed by preparation of the initial state, and as we show below, the formation dynamics redistributes the energy among the components of Eq. (21) in a fixed total energy.

\begin{figure}[htbp]
\centering
\includegraphics[width=\columnwidth]{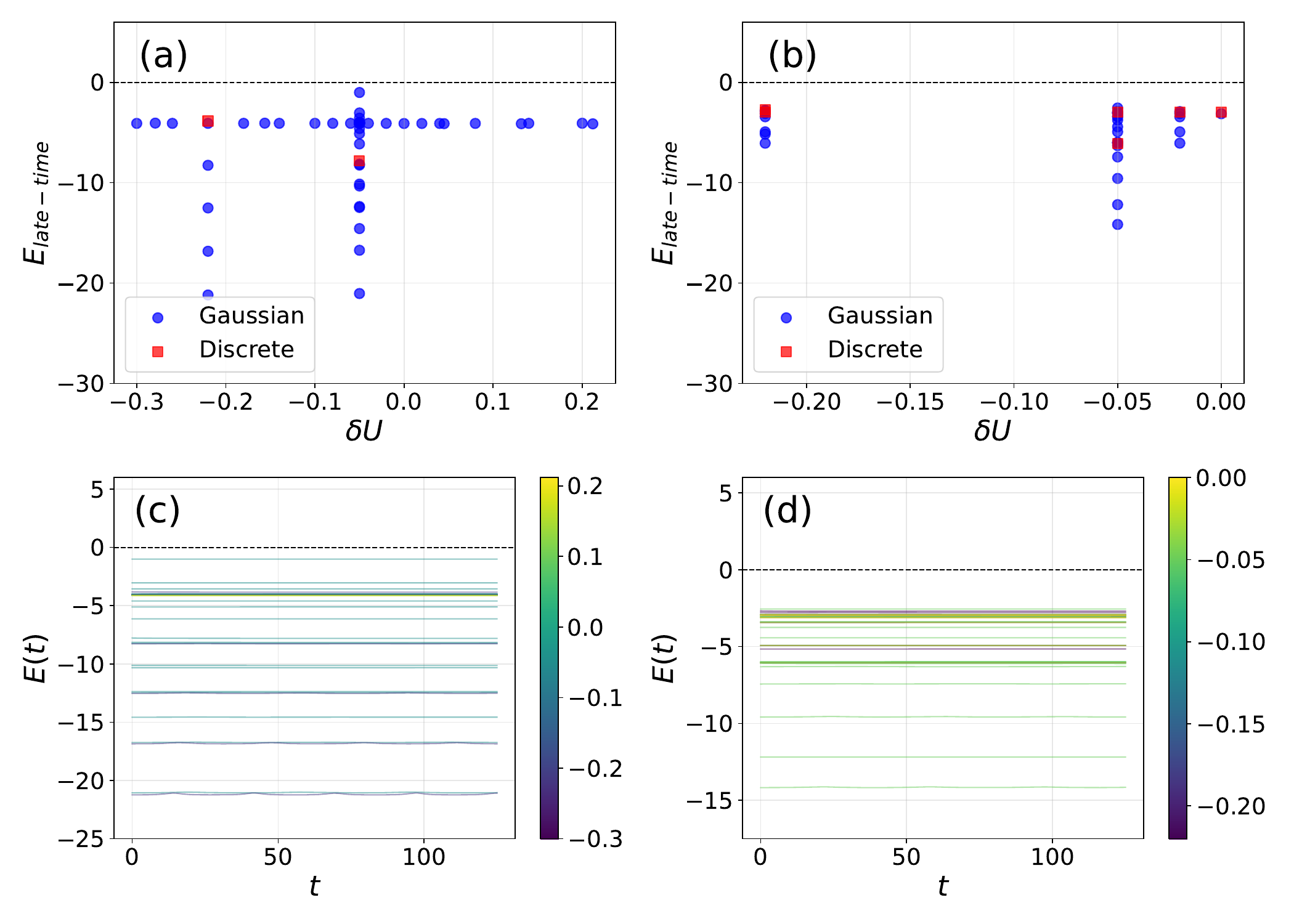}
\caption{Late-time total energy $E_{\text{late-time}}$ versus $\delta U$ for (a) homonuclear and (b) heteronuclear cases; the discrete bands correspond to the free-atom reference energies $-2(J_1N_1+J_2N_2)$ of the different particle-number families. Total energy time evolution $E(t)$ with color bars indicating $\delta U$ values for (c) homonuclear and (d) heteronuclear cases, demonstrating its conservation.}
\label{fig:late_time_energy}
\end{figure}
 
Figure \ref{fig:late_time_energy} provides the total energy value analysis of the simulated droplets. With the free-atom reference subtracted, the bindings are of order $10^{-1}$ in dimensionless units. For the subset of 45 formed droplets with $N_1 = N_2 = 1$, Gaussian initial conditions and no initial separation, which allows a controlled comparison, homonuclear and heteronuclear systems bind with statistically indistinguishable depths:
\begin{align}
E_{\text{bind}}^{\text{(homo)}} &= -0.082 \pm 0.019\\
E_{\text{bind}}^{\text{(hetero)}} &= -0.090 \pm 0.022.
\end{align}
Mass asymmetry is therefore not a substantial advantage to binding; it is, however, the controlling variable within the heteronuclear family: the binding deepens monotonically with the mass ratio, with a Pearson correlation of $r = -0.94$ between $E_{\text{bind}}$ and $m_2/m_1$. The deeper binding at larger mass ratio can be understood from the reduced kinetic energy cost when heavier atoms participate in the droplet: the kinetic energy scales as $\hbar^2/(2m)$ (on the lattice, this is equivalent to $J_2 = J_1\, m_1/m_2$), so replacing one component with heavier atoms reduces the kinetic energy penalty for localization. 
 
Because the total energy is conserved, no transient energy minimum exists, and $E_{\min}$ and $E_{\text{late-time}}$ coincide within the numerical accuracy of the integration scheme, for every run and both nuclear types. The transient compression that follows the preparation, and the subsequent collective oscillation, proceed instead as an exchange between the kinetic and interaction components of Eq. (\ref{energy-components}) at fixed total energy, as addressed below.
 
The particle number provides the strongest energetic handle in the data set. Along two families of homonuclear Gaussian co-localized equal-population runs with $N_1 = N_2 = N \in [1,5]$, one for $\delta U = -0.05$ and another for $\delta U = -0.22$, the binding energy follows a power law,
\begin{equation}
|E_{\text{bind}}| \simeq 0.08\, N^{1.6},
\end{equation}
with fitted exponents 1.60 and 1.66 in the two families, which is a markedly super-extensive scaling of the self-binding with atom number.
 
For the $N_1 = N_2 = 1$ Gaussian co-localized subset, the binding shows at most a weak dependence on the interaction strength: the Pearson correlation between $E_{\text{bind}}$ and $\delta U$ is $r = -0.24$, and $r = -0.27$ within the equilibrated subset; pooled over all formed droplets it almost vanishes ($r = 0.11$). This near-independence reflects the importance of quantum fluctuations in this regime: the LHY correction provides the dominant stabilization mechanism, so the binding depth is largely insensitive to the residual mean-field interaction $\delta U$, leading to robust droplet formation even when the mean-field attraction is weak, or, as in the $\delta U > 0$ runs, net repulsive. Indeed, decomposing the interaction energy at $t=125$ of the canonical equilibrated droplet ($\delta U = -0.05$) gives a mean-field interaction energy of $-0.004$ against an LHY energy of $-0.111$, and for a repulsive-side droplet ($\delta U = +0.20$) the mean-field interaction energy is net positive ($+0.012$) while the LHY energy, $-0.105$, supplies the entire attraction.
 
The preparation of the initial state affects the binding in the direction dictated by energy conservation: excess preparation energy cannot leave the system. Discrete initial conditions contain substantial high-momentum components (a manifestation of the uncertainty principle), and that energy is retained, with the formed discrete droplets having a median $E_{\text{bind}} = +0.04$, against $-0.084$ for Gaussian preparations. This results in a bound core coexisting with radiation that keeps the total energy above the free-atom reference. Initially separated configurations behave analogously: the separation is a form of potential energy that remains in the system after coalescence, giving a median $E_{\text{bind}} = -0.016$ for the 14 separated Gaussian droplets against $E_{\text{bind}} =-0.082$ for co-located population $N_1=N_2=1$. At the largest separations (20, 30, and 50 sites) the initial clouds are essentially disjoint, and the systems sit at the binding threshold itself: the $m_2/m_1=0.5$ heteronuclear pairs end marginally unbound ($E_{\text{bind}} \approx +0.017$) even though a droplet forms, while their $m_2/m_1 = 2$ counterparts ($E_{\text{bind}} \approx -0.024$) and the homonuclear droplets (average $E_{\text{bind}} = -0.0028 \pm 0.0033$) remain bound, which is the ordering expected from the mass-ratio dependence of the binding. Both trends are direct manifestations of the weak dissipation of one-dimensional dynamics.

\begin{figure}[htbp]
\centering
\includegraphics[width=\columnwidth]{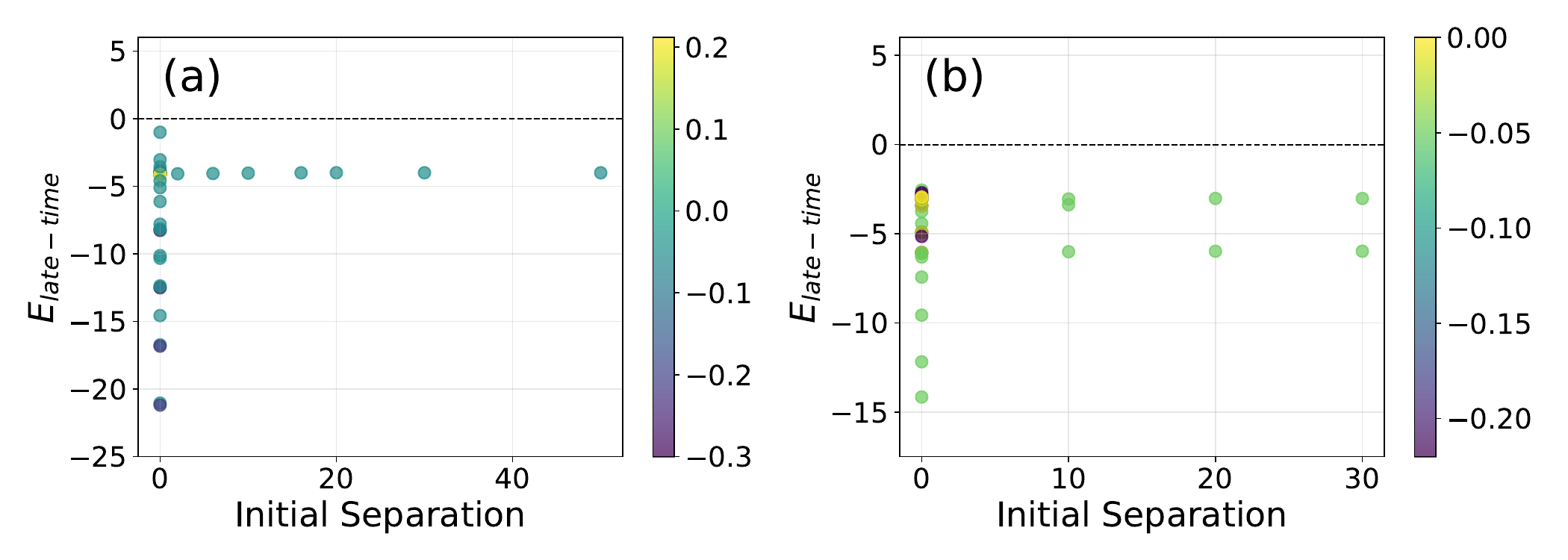}
\caption{Late-time total energy $E_{\text{late-time}}$ versus initial separation for (a) homonuclear and (b) heteronuclear droplets. The colored rule bar indicates the scale of $\delta U$.}
\label{fig:binding_sep}
\end{figure}

Figure \ref{fig:binding_sep} displays the total energy in the late-time as a function of the initial separation. Because the total energy is conserved and dominated by the reference of the free-atom of each family of decks, the data are organized into horizontal bands at $E_{\text{free}} = -2\,(J_1 N_1 + J_2 N_2)$: a single band at $-4$ for homonuclear systems [panel (a)] and two bands, at $-3$ and $-6$, for heteronuclear systems [panel (b)], corresponding to the two mass ratios simulated through $J_2 = J_1\, m_1/m_2$. Separate preparations, all with $N_1 = N_2 = 1$, lie in the band of their family at every separation: at the scale of the band structure, the dependence of the total energy on the initial separation is essentially indistinguishable because it is confined to the offset of each point from its band (the binding energy of order $10^{-1}$ analyzed above). That offset shrinks as the separation grows, since the interaction energy between the initial clouds is progressively lost and saturates once the clouds are essentially disjoint (separations $\gtrsim 4\sigma$), which is the band-level view of the threshold behavior quantified in the preceding paragraph. The vertical stacks at zero separation are the co-located families with larger particle numbers, whose deeper bands the panels also resolve.
 
The relationship between the binding and droplet structure is now consistent between nuclear types. For the $N_1=N_2=1$ Gaussian co-localized droplets, the binding shows a clear positive correlation with width for homonuclear ($r = 0.54$) and heteronuclear ($r = 0.77$) systems alike: wider droplets have shallower binding, as expected from the reduced density and weaker interactions.
 
As stated in the previous subsection, the equilibration criteria place conditions only on the shape dynamics. An independent energetic stability criterion is not meaningful here as the total energy is conserved to the level quoted above, which would make every run trivially energy stable. Therefore, equilibration in this system is a structural property referring to the damping of collective motion at fixed total energy. We return to this topic in Section \ref{sec:breathing-mode}.
 
The restriction to the equilibrated subset remains informative for the variance structure. Across all 93 formed droplets, the late-time binding energy spans two orders of magnitude ($CV = 1.56$), but this scatter is dominated by the particle-number and mass-ratio structure identified above rather than by transient dynamics: within the $N_1=N_2=1$ Gaussian co-localized subset the coefficient of variation is already $0.24$, and the equilibrated subset gives the following
\begin{align}
E_{\text{bind}}^{\text{(eq)}} &= -0.084 \pm 0.017,
\end{align}
with $CV = 0.20$. The apparent contamination of the ensemble statistics by non-equilibrated trajectories is therefore largely resolved into deterministic parameter dependence. The restricted analysis comes with the caveats already noted: the 15 equilibrated droplets all share Gaussian initial conditions, zero initial separation, and mass ratios in the narrow range $m_2/m_1 \in [1.0, 2.0]$ (13 homonuclear, 2 heteronuclear at $m_2/m_1 = 2$), and the trends involving discrete and separated preparations rest on small samples (8 and 14 formed runs, respectively) and should be regarded as tentative pending simulations extended substantially beyond the present horizon. We revisit these implications in the context of breathing-mode damping in Section \ref{sec:breathing-mode} and in the conclusions.

\begin{figure}[H]
\centering
\includegraphics[width=0.91\columnwidth]{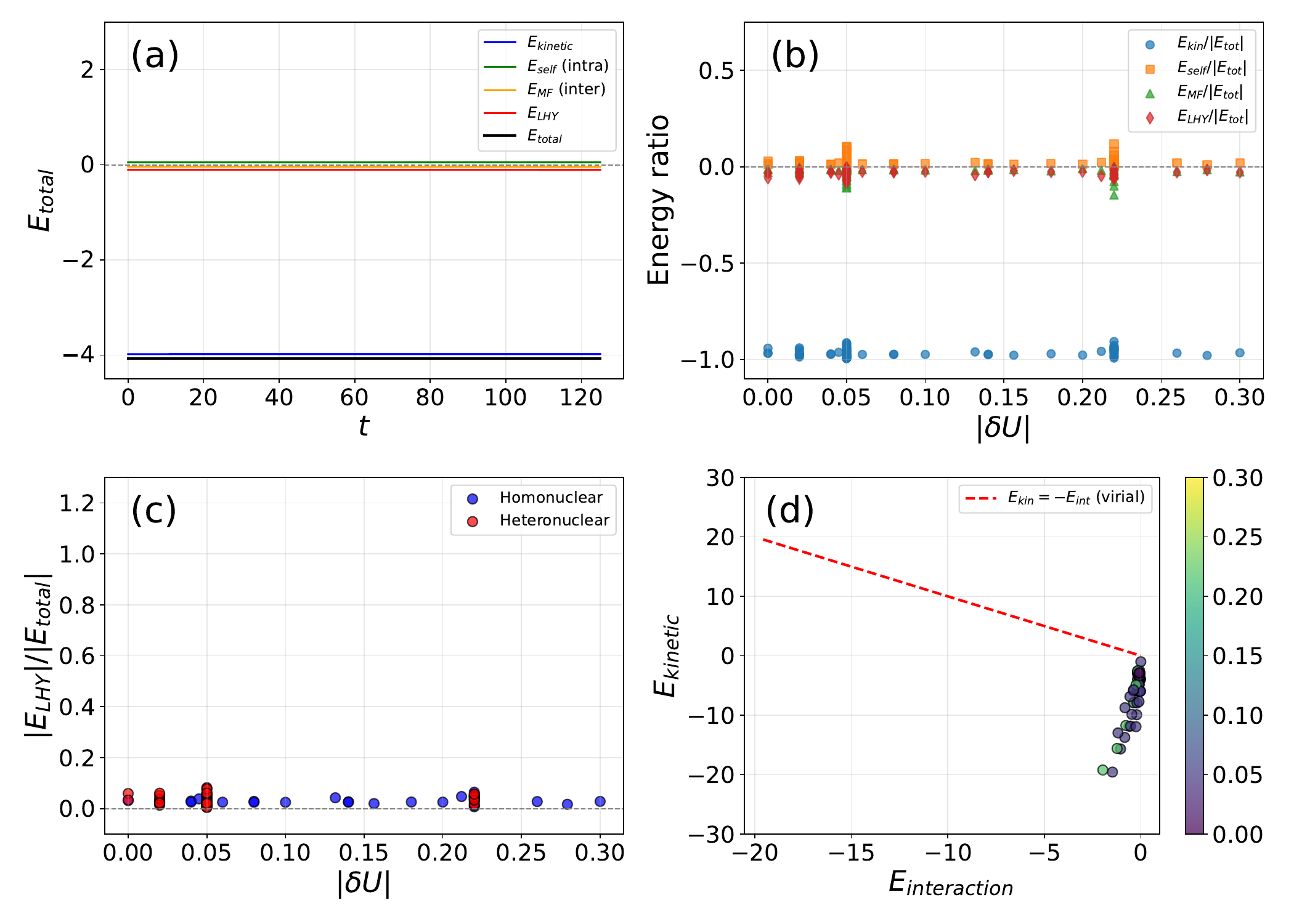}
\caption{Energy partitioning analysis: (a) Evolution of the energy components for a representative run, showing the exchange between kinetic and interaction energies at conserved total energy; (b) Energy ratios at late-time versus $|\delta U|$: the total is dominated by the kinetic band contribution, $E_{\text{kin}}/|E_{\text{total}}| \approx -0.97$; (c) LHY fraction of the late-time energy, $|E_{\text{LHY}}|/|E_{\text{total}}| \approx 0.03$: small in the total, yet dominant in the interaction budget and hence in the binding; (d) Kinetic energy (due to hopping) versus interaction energy (due to self-interaction, mean-field interaction and LHY correction) scatter plot. Colorbar indicates the values $\vert\delta U\vert$ of the interaction strength.}
\label{fig:energy_partition}
\end{figure}

 Figure \ref{fig:energy_partition} presents the analysis of energy partitioning, revealing the central result of this section. The total energy is dominated by the contribution of the kinetic band, which the bound and free configurations share; the binding itself is decided within the interaction budget, and there the dominance of the LHY term is complete: the components of the mean-field interaction energy within the species and between species nearly cancel each other, and its result is net repulsive for the few cases in which $\delta U > 0$, while the LHY energy supplies essentially all the attraction, as quantified above. This confirms that quantum fluctuations constitute the essential binding mechanism in 1D droplets, not merely a perturbative correction: self-binding is a small residual effect carried out by $E_{\text{LHY}}$ against the kinetic background of $-2(J_1N_1+J_2N_2)$.

\begin{figure}[H]
\centering
\includegraphics[width=\columnwidth]{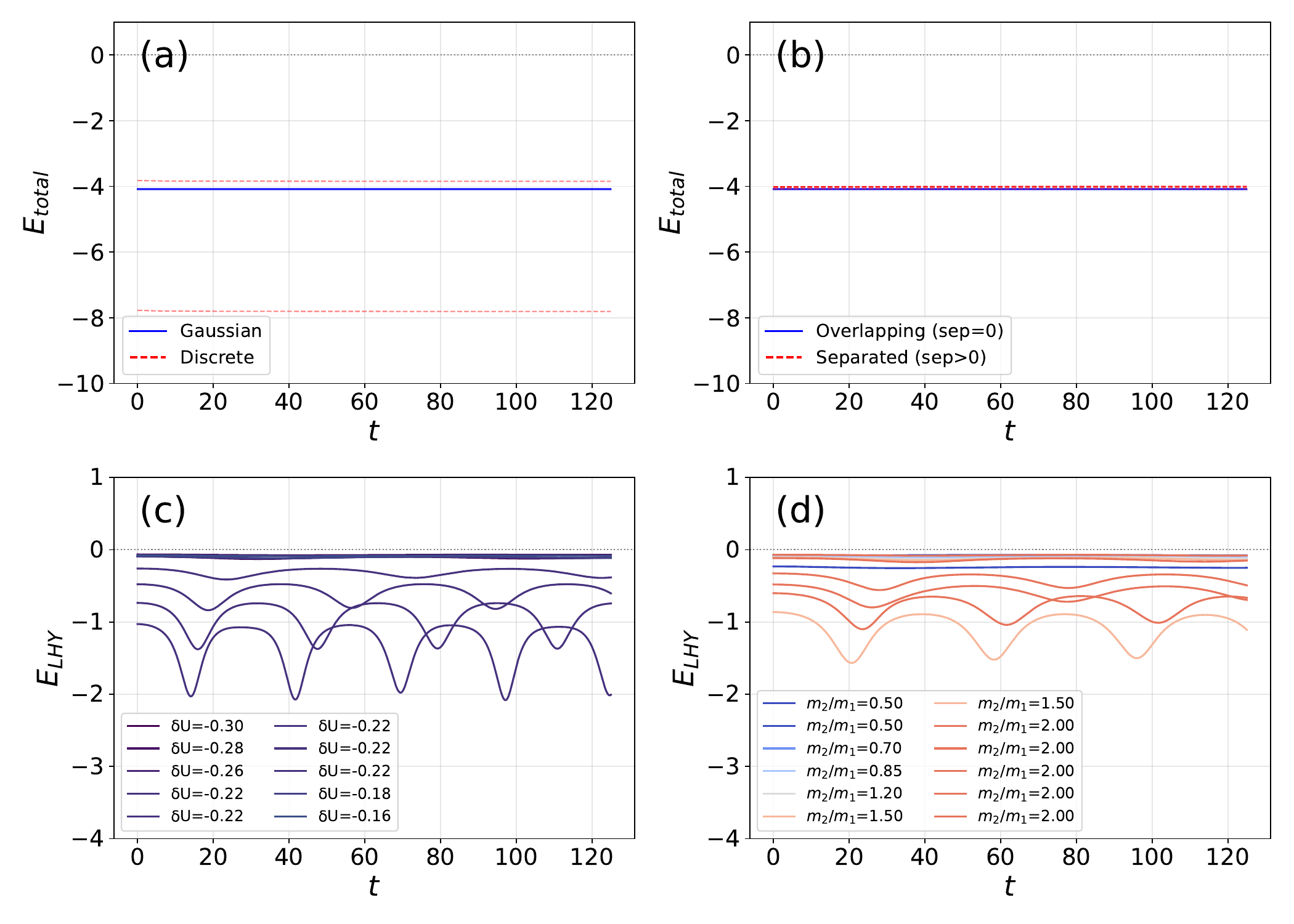}
\caption{Energy dynamics comparison. Evolution of the total energy for (a) Gaussian and discrete initial conditions, and for (b) initially co-localized and separated configurations. Evolution of the LHY energy for (c) various homonuclear runs, colored by $\delta U$, and for (d) various heteronuclear runs, colored by mass ratio. We notice the breathing oscillations in time.}
\label{fig:energy_dynamics}
\end{figure}

Figure \ref{fig:energy_dynamics} compares the evolution of energy between nuclear configurations and initial conditions. Panel (a) shows conserved total energies for both preparation classes, with the discrete runs displaced upward by their retained quench energy; panels (c) and (d) show the LHY energy exchanging with the kinetic component as the droplet breathes.

The low equilibration rate, with 16.1\% of the formed droplets width stable, reflects fundamental constraints of one-dimensional dynamics. The near-integrability of 1D systems limits thermalization channels, causing droplets to retain collective excitations over timescales much longer than typical experimental windows.
 
\subsection{Localization Measures: Inverse Participation Ratio and Shannon's Entropy}\label{sec:localization-measures}

We employ two complementary measures to quantify the degree of spatial localization of the droplets on the lattice: the inverse participation ratio (IPR) and Shannon's entropy. 

The inverse participation ratio is defined as 
\begin{equation}
    \text{IPR} = \left(\frac{\sum_i \vert\psi_i\vert^4}{\left(\sum_i\vert\psi_i\vert^2\right)^2}\right)^{-1},
\end{equation}
summed over all lattice sites. It captures how many sites are effectively occupied by the wave function of the system: IPR $\approx 1$ indicates extreme localization; essentially, only one site is occupied by the condensate(s). On the other hand, IPR $\approx N_{\text{sites}}$ indicates complete delocalization; all sites are being equally occupied. For a Gaussian profile of width $\sigma$, IPR $\propto \sigma$.

Shannon's entropy is defined as

\begin{equation}
    S = -\sum_i p_i\ln p_i,\quad p_i = \frac{|\psi_i|^2}{\sum_j |\psi_j|^2}
\end{equation}
It is another measure of how ``spreads out" the probability distribution of the system. Higher values indicate more delocalization. 

\begin{figure}[htbp]
\centering
\includegraphics[width=\columnwidth]{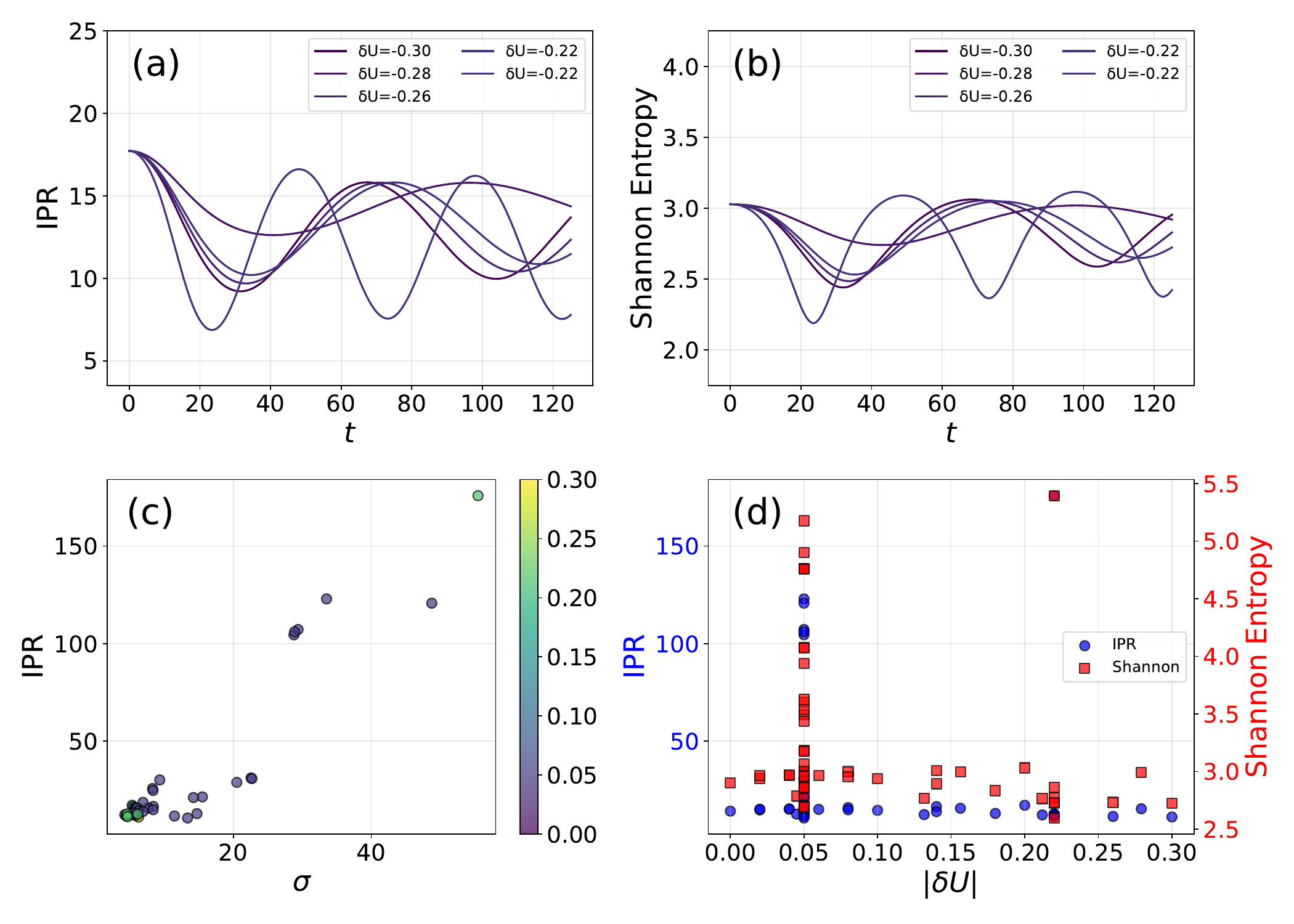}
\caption{Localization measures for homonuclear droplets: (a) IPR evolution versus time for different most negative $\delta U$ values; (b) Same for Shannon entropy evolution; (c) IPR versus width showing a strong correlation of 0.933; (d) Localization measures IPR and Shannon averages versus $|\delta U|$ for the final 30\% of the simulation.}
\label{fig:localization-homo}
\end{figure}

Figure \ref{fig:localization-homo} presents the localization analysis for homonuclear droplets. To avoid data contamination, panels \ref{fig:localization-homo}(a) and \ref{fig:localization-homo}(b) only show at most two initially co-localized Gaussian states for the four most negative $\delta U$. As panels \ref{fig:localization-homo}(a) and \ref{fig:localization-homo}(b) show, both the IPR and Shannon's entropy display breathing oscillations mirroring width dynamics. We find that the IPR value and, therefore, the effective localization, is contingent on the interaction strength $\delta U$, though not completely determined by such a parameter, as several droplets that share the same $\delta U$ value present different IPR and Shannon dynamics. Importantly, IPR shows no secular drift -- it oscillates around a stable mean, confirming that the droplets maintain a self-bound character without dispersing or collapsing. In particular, panel \ref{fig:localization-homo}(c) presents a correlation of 0.933 between IPR and width, which is very strong, indicating that, for homonuclear droplets, IPR serves as a useful proxy for droplet size independent of the specific profile shape. 

\begin{figure}[htbp]
\centering
\includegraphics[width=\columnwidth]{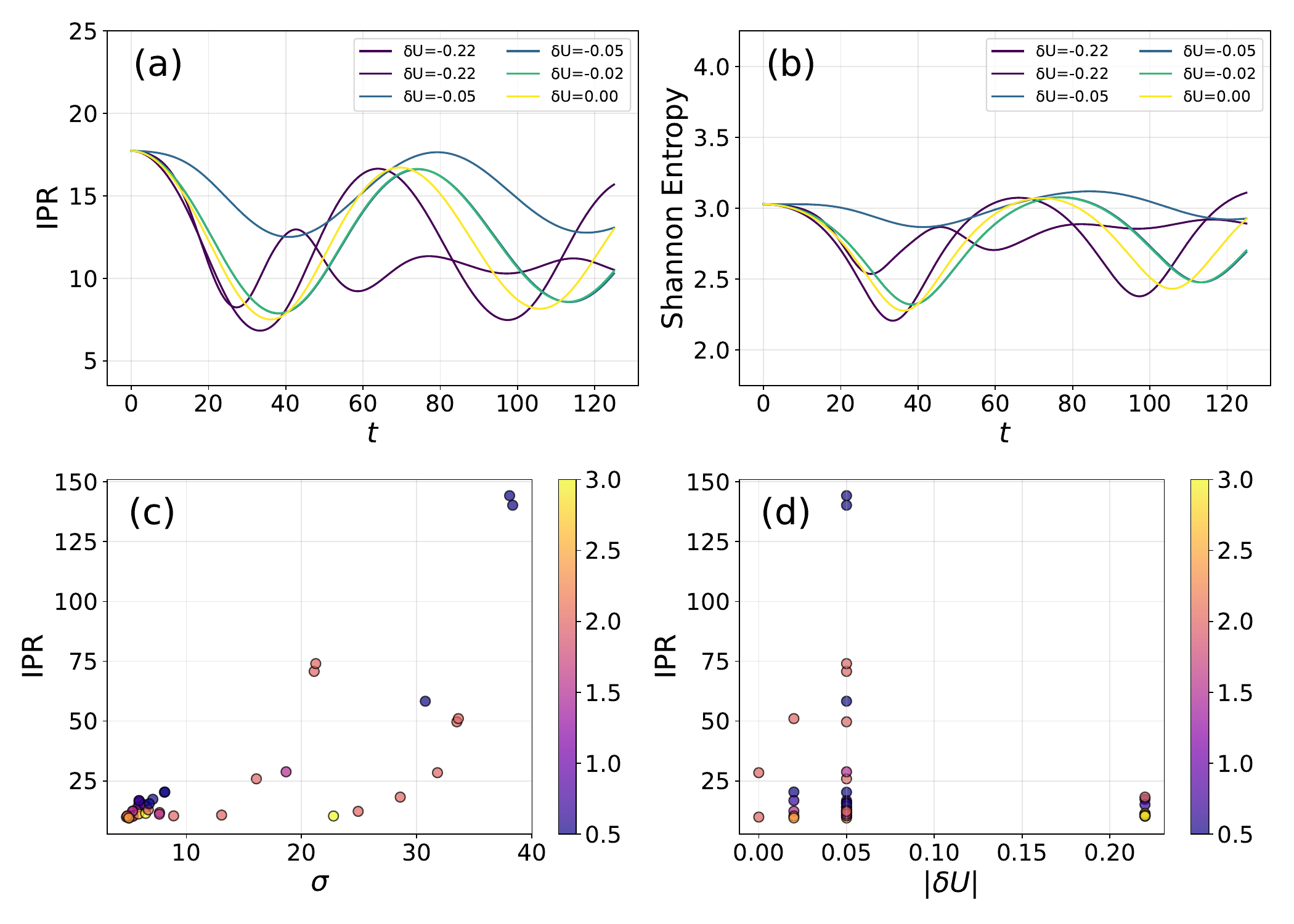}
\caption{Localization measures for heteronuclear droplets: (a) IPR evolution versus time for different $\delta U$ most negative values; (b) Same for the Shannon entropy evolution; (c) IPR versus width showing a correlation of 0.750; (d) Localization IPR measure averages versus $|\delta U|$ for the final 30\% of the simulation.}
\label{fig:localization-hetero}
\end{figure}

Figure \ref{fig:localization-hetero}, on the other hand, presents the localization measures for heteronuclear droplets. Once again, we only present at most two curves for initially co-localized Gaussian states for the four most negative values of $\delta U$, now only for a $m_2/m_1=2$ mass ratio. 

The most interesting conclusion we can reach, from the correlation of 0.750 between the inverse participation ratio and the width in panel \ref{fig:localization-hetero}(c), is that IPR serves as a decent albeit weak proxy for droplet size when there is mass-imbalance between atomic species. Data is more scattered now.

The peak-to-edge ratio $\mathcal{L} = n_{\text{peak}}/n_{\text{edge}}$, which is one of the metrics used to evaluate droplet formation, shows a great variation between nuclear types: for homonuclear droplets, we have a median $\mathcal{L}\sim 10^7$, while for heteronuclear droplets, we find a median $\mathcal{L}\sim 2.7\times 10^5$. This factor of $\sim 37$ between cases reflects the more compact density profiles enabled by symmetric mixtures in homonuclear systems, where both components contribute equally to the central peak density.

It should be noted that initial conditions can have a dramatic effect on the final spatial localization. Gaussian initial states are correlated with compact droplet formations with a median $\mathcal{L} \approx 8\times10^6$, whereas discrete initial configurations lead to broad structures with a median $\mathcal{L}\approx5.7\times10^2$. This may seem paradoxical, as the more localized discrete conditions result in more delocalized states, but we believe it is a manifestation of the uncertainty principle: the extreme initial localization (with a very small spatial uncertainty) must have a correspondingly large momentum uncertainty. This range of momentum states, and its accompanying kinetic energy, must be accommodated as the system relaxes to an equilibrium stabilized by the LHY correction. However, the sheer non-physicality of delta-like states limits the sample size, as many simulations simply do not lead to droplet formation in the first place.

\subsection{Component Overlap and Coalescence Dynamics}\label{sec:overlap}

For two-component droplets to achieve their maximal binding energy, the two species must overlap spatially. We characterize this through the normalized overlap integral:
\begin{equation}
\mathcal{O}(t) = \frac{\int dx\, |\psi(x,t)|^2 |\phi(x,t)|^2}{\sqrt{\int dx\, |\psi|^4 \int dx\, |\phi|^4}}
\label{eq:overlap}
\end{equation}
which equals unity for perfectly overlapping distributions and zero for completely separated components.

\begin{figure}[htbp]
\centering
\includegraphics[width=\columnwidth]{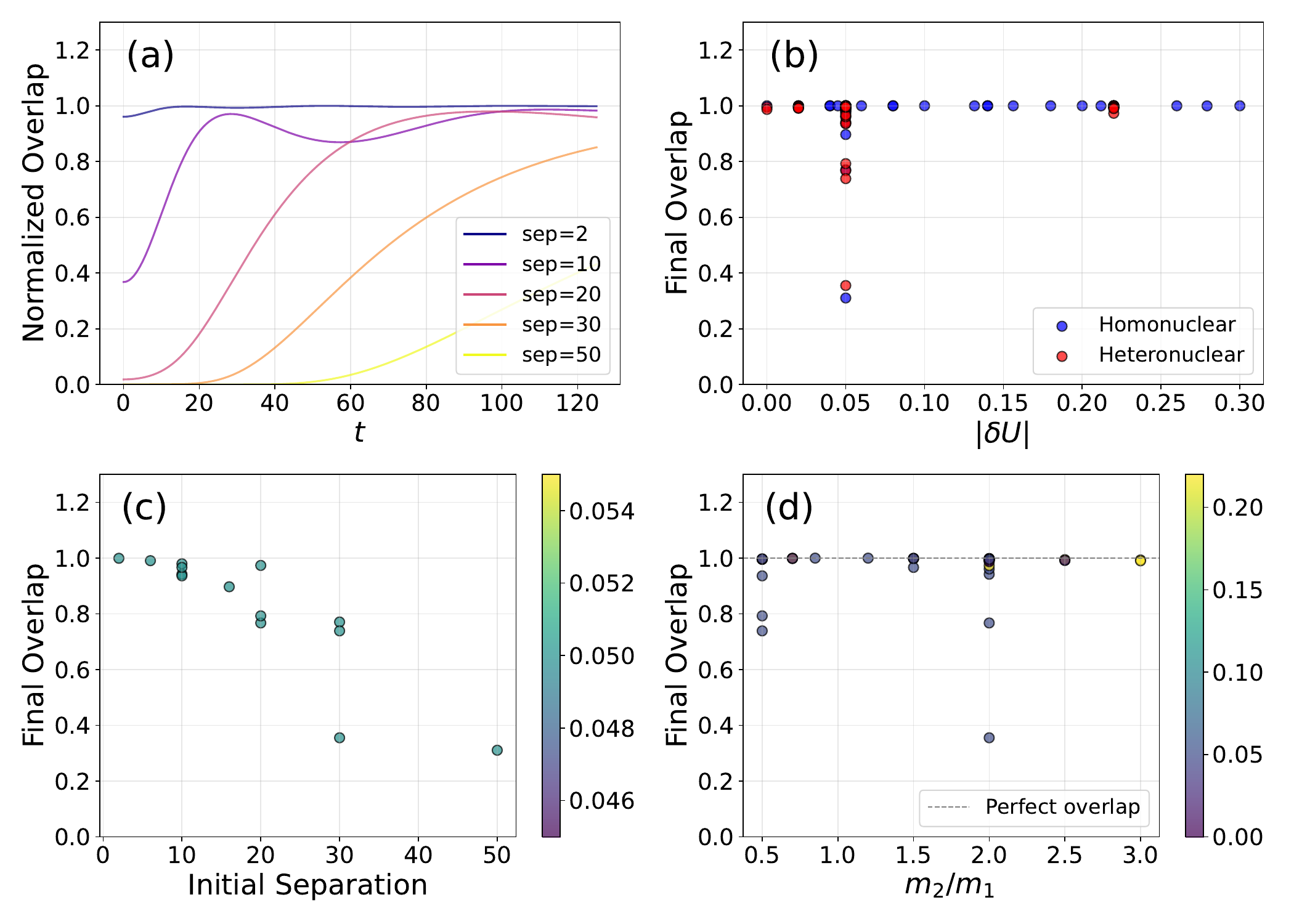}
\caption{Component overlap and miscibility: (a) Overlap evolution for representative homonuclear runs with different initial separations; (b) Final overlap for both classes versus $|\delta U|$; (c) Final overlap versus initial separation; (d) Final overlap versus mass ratio. Color bars on panels (c) and (d) represent values of  $\vert \delta U\vert$.}
\label{fig:overlap}
\end{figure}

Figure \ref{fig:overlap} presents the overlap analysis. The final (or late-time) overlap is computed as the average of the instantaneous overlap over the final 30\% of the simulation, to smooth out irrelevant oscillations. By analyzing panel \ref{fig:overlap}(a), we can see that systems with separated initial conditions have overlap increase from near zero to near unity over timescales that depend on the initial separation. The coalescence dynamics in most cases shows clear sigmoidal behaviors, with a slow initial approach, a rapid merging, and stabilization at high values of $\mathcal{O}$. Panels \ref{fig:overlap}(b) - \ref{fig:overlap} (d) show the late-time average overlap across all 93 formed droplets; for the 78 un-equilibrated runs, this average reflects an ongoing relaxation rather than a stationary value, which contributes to the scatter at fixed $\vert \delta U\vert$. For an initial separation of 2 sites, we achieve $\mathcal{O} \approx 0.98$ by $t \approx 10$. For condensates initially 10 sites apart, we reach $\mathcal{O} \approx 0.9$ by $t \approx 20$. The oscillations displayed over time indicate that the dynamics of capture of initially separated condensates is not necessarily monotonic. 

The panel \ref{fig:overlap}(b) shows that most of the values of $\vert\delta U\vert$ used are sufficient to warrant near-perfect coalescence ($\mathcal{O}>0.95$) for both homonuclear and heteronuclear systems. The exceptions occur for very small values of $\vert\delta U\vert$, where the attraction between components provides insufficient driving force for the capture of both condensates into a single structure to occur. We can see from \ref{fig:overlap}(c) that there is a systematic decrease in the final overlap reached with increased initial separation: zero separation maintains $\mathcal{O} \approx 1.0$; separations up to 20 sites achieve $\mathcal{O}>0.8$; beyond 30 sites, overlap drops to below $\mathcal{O}\sim 0.4$. This relationship is largely independent of $|\delta U|$, being dominated by geometric constraints and limitations of the simulated time scale. Remarkably, \ref{fig:overlap}(d) shows that the final overlap has essentially no dependence on the mass ratio. Almost all heteronuclear mixtures, for $m_2/m_1$ varying between $0.5$ and $3.0$, reach $\mathcal{O}>0.95$, showing that the coalescence dynamics is robust against kinetic energy differences.

The high final overlaps indicate that the droplet state is miscible - both components occupy the same region rather than phase-separating. This is expected, as the attractive inter-component interaction ($U_{12} < 0$) favors overlap, and the LHY correction further stabilizes the mixed configuration.

\subsection{Breathing Mode Analysis}\label{sec:breathing-mode}

In the context of droplet physics, the term ``breathing" refers to coherent oscillations in the width of the core fraction of the system that is identified as the droplet. It is the lowest-lying form of collective excitation for our system. In 1D, it is important to recall that the overall topology of the corresponding phase space is quite restricted, and hence there are limited mechanisms for dissipation of energy and momentum. Nevertheless, the dynamics of this Bose mixture, especially at the droplet-forming threshold, can still be very rich.

Once a droplet has formed, the simulated width $w(t)$
does not reach a strictly stationary value within our simulation horizon; rather, it relaxes about an equilibrium configuration while continuing to oscillate. We model the late-time width as
\begin{equation}
w(t) \approx w_{\text{eq}}(t) + A\,e^{-\gamma t}\cos(\omega_B\, t + \phi) + \eta(t),
\label{eq:width_decomposition}
\end{equation}
where $w_{\text{eq}}(t)$ is the slowly varying equilibrium width that captures the residual relaxation towards the asymptotic state, $A$, $\gamma$, $\omega_B$, and $\phi$ are respectively the amplitude, damping rate, angular frequency, and phase of the breathing oscillation, and $\eta(t)$ represents the numerical error introduced from the Crank-Nicolson integrator and the discrete grid. The breathing frequency $\omega_B$ and the damping rate $\gamma$ together characterize the dominant collective excitation of the droplet: $\omega_B$ sets the timescale on which the droplet rhythmically contracts and expands about its equilibrium width, while $\gamma$ sets the timescale on which this oscillation decays as energy flows into other degrees of freedom. Equivalently, the pair can be packaged as a complex frequency $\Omega = \omega_B - i\gamma$, with the real part describing the oscillation and the imaginary part describing its decay; the dimensionless ratio $\gamma/\omega_B$ then quantifies the fraction of energy lost per radian of breathing and serves as an inverse quality factor for the mode. We extract $\omega_B$ and $\gamma$ separately, using complementary signal-processing techniques as described below. The fits are performed in the detrended series $w(t) - w_{\text{eq}}(t)$, with $w_{\text{eq}}(t)$ approximated by a low-degree polynomial fit (linear for frequency analysis, quadratic for damping analysis) so that the slow drift contribution is largely removed before the oscillation is analyzed.

The damping analysis relies on extracting the time-dependent amplitude envelope of the detrended width signal, which is best accomplished through the analytic-signal representation. Given the real-valued detrended signal $s(t) = w(t) - w_{\text{eq}}(t)$, we form the complex-valued analytic signal $z(t) = s(t) + i \mathcal{H}s$, where $\mathcal{H}$ is the Hilbert transform, the linear operator that produces the phase-shifted partner $90^{\circ}$ of s \cite{oppenheim2005discrete}. For an undamped sinusoid $s(t) = A\text{cos}(\omega_B t)$, this construction yields $z(t) = A e^{i\omega_B t}$, so the modulus $\vert z(t)\vert = A$ recovers the amplitude with the oscillation removed. For the damped case relevant here, $\vert z(t)\vert = A e^{-\gamma t}$, and the damping rate is recovered from a linear fit of $\text{log} \vert z(t)\vert$ versus t. This procedure is preferable to peak-finding methods or to taking $\vert s(t)\vert$ directly, because the analytic-signal envelope is smooth and sampled at the same density as the original time series, with no half-rectification artifacts and no need to interpolate between discrete oscillation maxima. In practice, we compute the analytic signal numerically using the standard FFT-based implementation, in which negative-frequency Fourier coefficients of $s(t)$ are zeroed, positive-frequency coefficients are doubled, and the inverse transform is taken to yield $z(t)$.

\begin{figure}[htbp]
\centering
\includegraphics[width=\columnwidth]{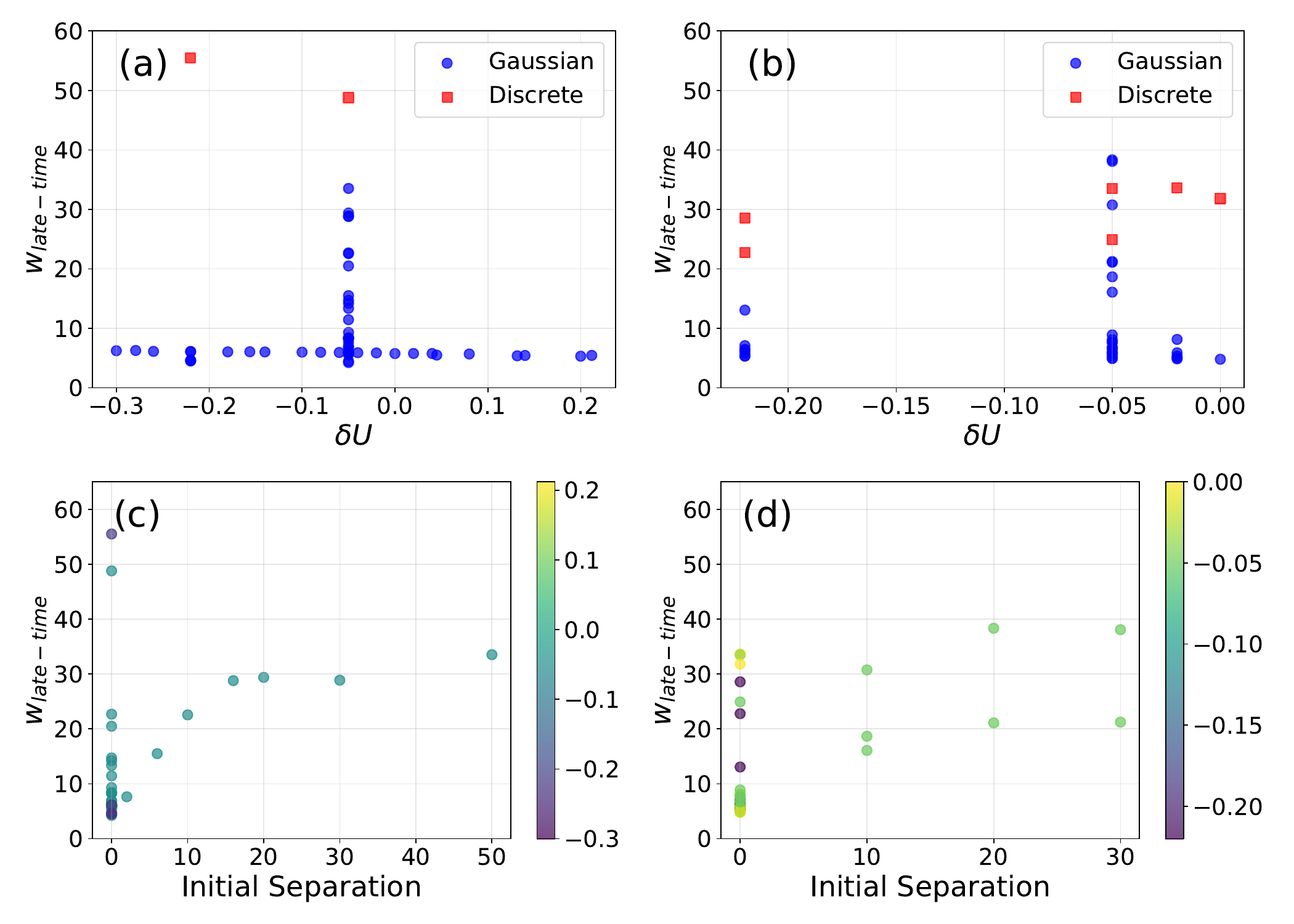}
\caption{Late-time width $w_{\text{late-time}}$ versus $\delta U$ for (a) homonuclear and (b) heteronuclear cases. Late-time width $w_{\text{late-time}}$ versus initial separation for (c) homonuclear and (d) heteronuclear droplets. Both colored ruling bars represent $\delta U$ values.}
\label{fig:width}
\end{figure}

Figure \ref{fig:width} presents the late-time width analysis (the average width over the final 30\% of the simulation), contrasting homonuclear systems (panels (a) and (c)) and heteronuclear systems (panels (b) and (d)).

It is clear that both homonuclear and heteronuclear droplets exhibit an interesting bimodal distribution of late-time widths, with most forming compact structures and a few more extended outliers. These cases of larger widths largely represent either discrete initial conditions or initial separation, for which the component clouds merge into an elongated formation rather than a compact droplet. In particular, note that the dependence on $\delta U$ is generally weak: compact formations that represent one of these modes are essentially independent of the interaction strength, while wider structures do not obey a clear trend. Once again we see the formation spike around $\delta U = -0.05$, which was the canonical interaction strength against which perturbations were measured. 

\begin{figure}[h]
\centering
\includegraphics[width=\columnwidth]{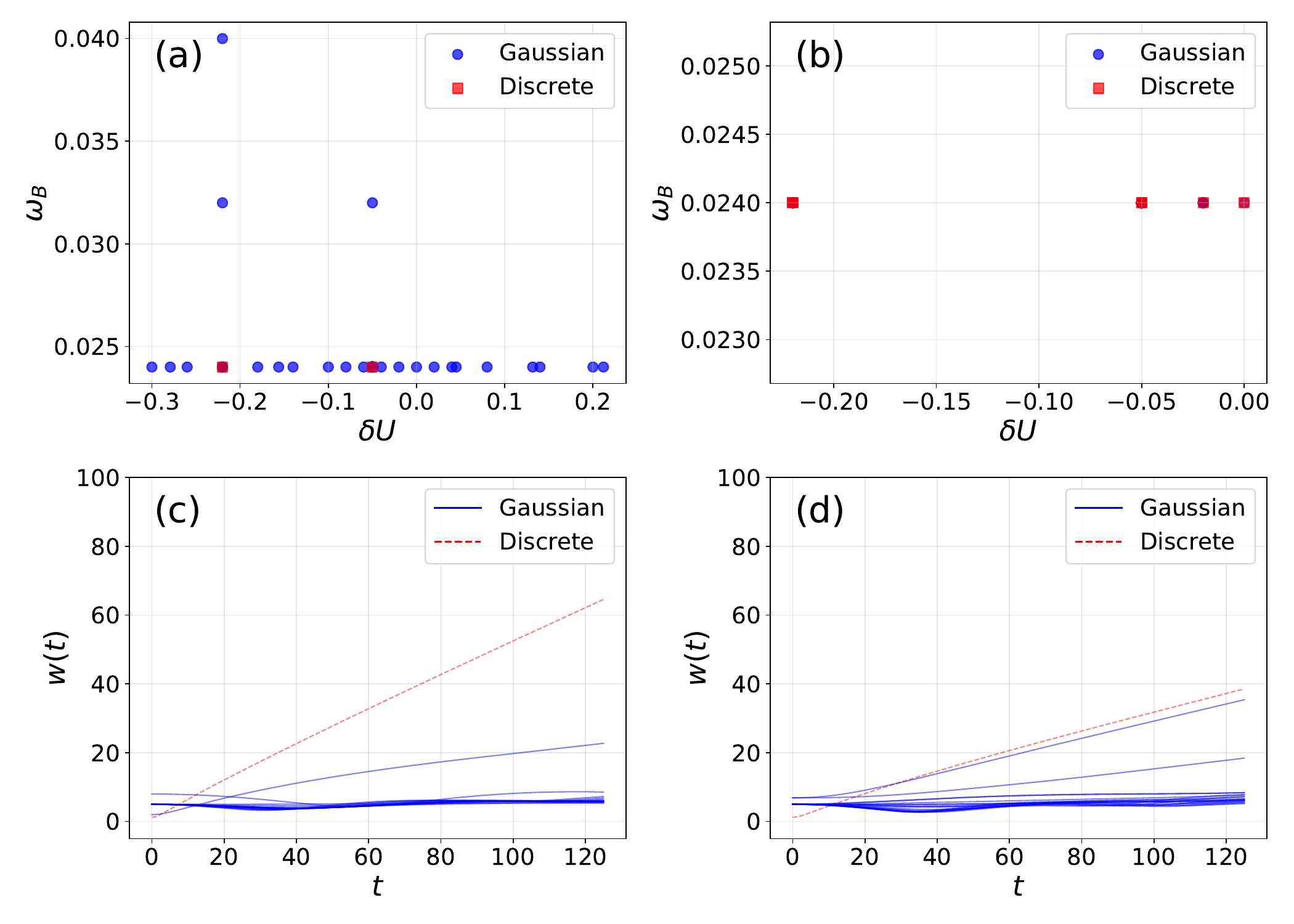}
\caption{Breathing mode analysis using Gaussian and discrete shapes. Breathing frequency $\omega_B$ versus $\delta U$ for (a) homonuclear and (b) heteronuclear cases. Width time series $w(t)$ for (c) homonuclear and (d) heteronuclear droplets.}
\label{fig:breathing}
\end{figure}

Figure \ref{fig:breathing} presents the analysis of the breathing mode. As stated above, we extract the breathing frequency using the Fast Fourier Transform (FFT) analysis of the time series of width $w(t)$ after linear detrending \cite{oppenheim2005discrete}. To avoid spurious detection of slow drift, we impose a frequency floor $f_{\text{min}} = 2/T$, where $T$ is the period of oscillation, thereby requiring at least 2 complete oscillations. Peak detection requires a Signal-to-Noise Ratio (SNR) greater than 10 relative to the spectral mean. We obtain valid breathing frequencies for all 93 simulations classified as having formed a droplet. 

Before reporting the extracted frequencies, it is important to make explicit the resolution limit imposed by the simulation horizon. A discrete Fourier transform of a signal of duration $t_{\text{max}}$ samples the frequency axis on a uniform grid with spacing $\Delta f = 1/t_{\text{max}}$, and the peak-finding step in our pipeline returns the center of the bin of whichever bin captures the largest spectral power. With $t_{\text{max}} = 125$ in dimensionless units, the spacing of the bin is $\Delta f = 0.008$, and the lowest bin admitted by the frequency floor $f_{\min} = 2/t_{\text{max}} = 0.016$
is the third bin \cite{press2007numerical}, centered at $\omega_B = 0.024$. The width of this bin spans the interval [0.020, 0.028], and any genuine breathing frequency falling anywhere in this range is reported as $\omega_B = 0.024$ by construction. This coarse grid reflects an underlying physical limitation rather than a numerical choice: the breathing period at $\omega_B \approx 0.024$ is $t_B = 1/\omega_B \approx 41.7$, so each simulation contains only complete breathing cycles $t_{\text{max}}/t_B \approx 3$. Three cycles is sufficient to detect that an oscillation is present and to localize its frequency to within a fraction of a bin, but not to distinguish frequencies that differ by less than approximately $\Delta f / \omega_B \approx 30\%$. Consequently, the values reported below should be read as bin-center estimates with a resolution-limited uncertainty
$\sigma_{\omega_B} \sim 0.004$. If the data presented in the plots below indicate independence of $\omega_B$ from the system parameters, the reader should understand that as a constraint on parameter dependence at the level of~30\%. A finer determination would require simulations that extend substantially beyond $t_{\text{max}} = 125$, which is  computationally very expensive, as mentioned above.

Of the 93 cases, 90 of these have a measured frequency that falls in the lowest admitted FFT bin, corresponding to $\omega_B = 0.024 \pm 0.004$, including all 39 heteronuclear droplets and 51 (out of 54) homonuclear droplets. Two homonuclear simulations report values in the next bin ($\omega_B = 0.032$) and one homonuclear  simulation in the bin above that ($\omega_B = 0.040$). Given the resolution limit, we do not interpret these as physically distinct breathing modes but rather as cases where the spectral peak fell on the upper side of a bin boundary. Further inspection of these outliers reveals that two of these three outliers are weakly-formed configurations with strong attractive interaction ($\delta U = -0.22$) with atypical particle numbers ($N_1 = N_2 \in \{4, 5\}$) and large breathing amplitudes ($A \gtrsim 0.4$). The third outlier, on the contrary, is a run under canonical conditions ($\delta U = -0.05$, $N_1=N_2=1$) with a modest breathing amplitude ($A \approx 0.17$); its assignment to a higher bin most likely reflects the spectral peak falling on the boundary between bins 3 and 4 rather than any distinct physical mechanism. We interpret all three outliers as artifacts of the bin grid rather than as evidence of distinct physical modes. Within the dominant population, the absence of detectable bin-to-bin variation is consistent with the breathing frequency being independent of $\delta U$, of the mass ratio $m_2/m_1$, and of the preparation of the initial-state, at the level of resolution accessible in our analysis $\sim 30\%$.

Panels \ref{fig:breathing}(c) and \ref{fig:breathing}(d) show that Gaussian initial conditions lead to clean sinusoidal oscillations without perceptible signs of amplitude decay or discontinuous dynamics over the simulated period, while discrete initial conditions have more complex dynamics, exhibiting superimposed fluctuations unrelated to droplet dynamics or monotonically increasing width, which represent cases of failed droplet formation.

\begin{figure}[htbp]
\centering
\includegraphics[width=\columnwidth]{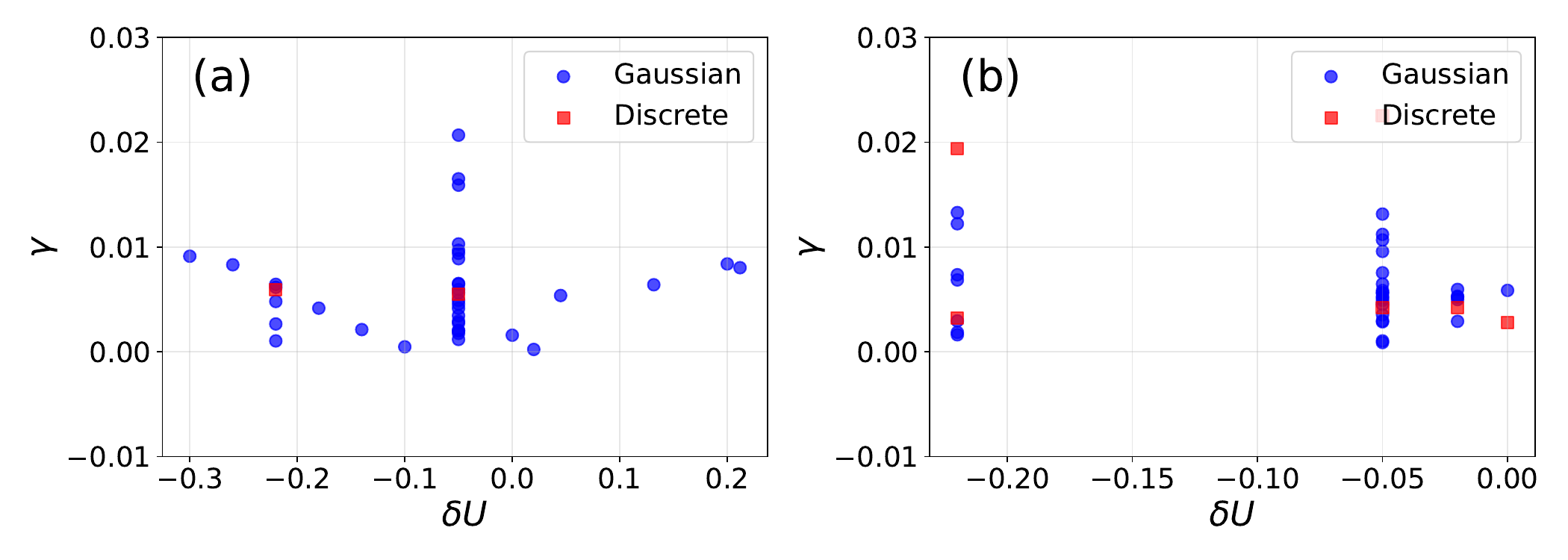}
\caption{Damping rate $\gamma$ versus $\delta U$ for (a) homonuclear and (b) heteronuclear droplets, for Gaussian and discrete cases.}
\label{fig:damping}
\end{figure}

Figure \ref{fig:damping} presents the damping rate analysis for homonuclear and heteronuclear droplets. As already stated, we computed the damping rate $\gamma$ from the envelope of detrended width oscillations using the Hilbert transform analysis. 

Damping of breathing oscillations proceeds through coupling to other degrees of freedom and eventual thermalization. We extract valid damping rates for 76 of the 93 droplets formed. The extraction of the damping rate failed for 17 runs, predominantly homonuclear configurations (15/17). Analysis reveals that these failures correlate with low breathing amplitude: runs with invalid damping exhibit mean amplitude $0.096 \pm 0.070$, compared to $0.238 \pm 0.133$ for successful extractions. The Hilbert transform envelope method requires measurable amplitude decay to fit an exponential; when oscillations are small (amplitude $\lesssim 10\%$), the envelope becomes essentially flat, yielding fitted decay rates indistinguishable from zero or slightly negative. Indeed, 6 of the 17 failed runs are classified as equilibrated, where stable width by definition precludes observable damping. Rather than report potentially spurious values, we conservatively exclude these runs from damping statistics. This limitation does not affect our physical conclusions, as the failed runs represent droplets that have reached equilibrium or exhibit oscillations too weak for reliable damping measurement.

The overall damping rate for these valid 76 simulations is found to be $\gamma = 0.00622 \pm 0.00458$ in dimensionless units, which corresponds to characteristic damping timescales of the 76 simulations.
\begin{equation}
    \tau_{\text{homonuclear}} = 1/\gamma_{\text{homonuclear}} \approx 167.7 ,
\end{equation}
\begin{equation}
    \tau_{\text{heteronuclear}} = 1/\gamma_{\text{heteronuclear}}  \approx 154.3 .
\end{equation}
Both timescales substantially exceed the simulated horizon $t_{\text{max}} = 125$, providing a quantitative explanation for why most droplets do not equilibrate within our simulations: the breathing mode simply does not have time to decay. Equivalently, in inverse-quality-factor form $Q^{-1} = \gamma/\omega_B \approx 0.26$, indicating that the breathing oscillation loses on the order of a quarter of its energy per radian, which is a moderately damped but long-lived collective excitation by the standards of one-dimensional dynamics. The theoretical expectation in 1D would be $\gamma \to 0$, as breathing modes lack effective coupling channels for momentum transfer in a strictly one-dimensional geometry. The non-vanishing $\gamma$ extracted here likely reflects a genuine but weak coupling to the radiation continuum at the droplet boundary, and a numerical contribution from the discrete lattice and finite Crank-Nicolson time step, whose relative magnitudes cannot be disentangled from the present data. In any case, $\tau_{\text{droplet}} \gg t_{\text{max}}$.

Another relevant characterization is the breathing amplitude, the width fraction that actually varies during the final 30\% of the simulation. We compute it from the width time series $w (t)$ as 
\[
\langle A\rangle = \frac{\text{max}(w_{\text{final 30\%}}) - \text{min}(w_{\text{final 30\%}})}{\langle w_{\text{final 30\%}}\rangle} \,.
\]
The mean amplitude fraction was $\langle A\rangle = 0.212\pm 0.135$, with $\langle A_{\text{homo}}\rangle = 0.191 \pm 0.143$ and $\langle A_{\text{hetero}}\rangle = 0.241 \pm 0.118$. This higher mean amplitude fraction for oscillations of heteronuclear droplets may reflect the additional degree of freedom of these kinds of systems: the relative motion between the two components with different masses, which may store oscillatory energy.

\subsection{Density Profile Characterization}\label{sec:density-profile}

\begin{figure}
\centering
\includegraphics[width=\columnwidth]{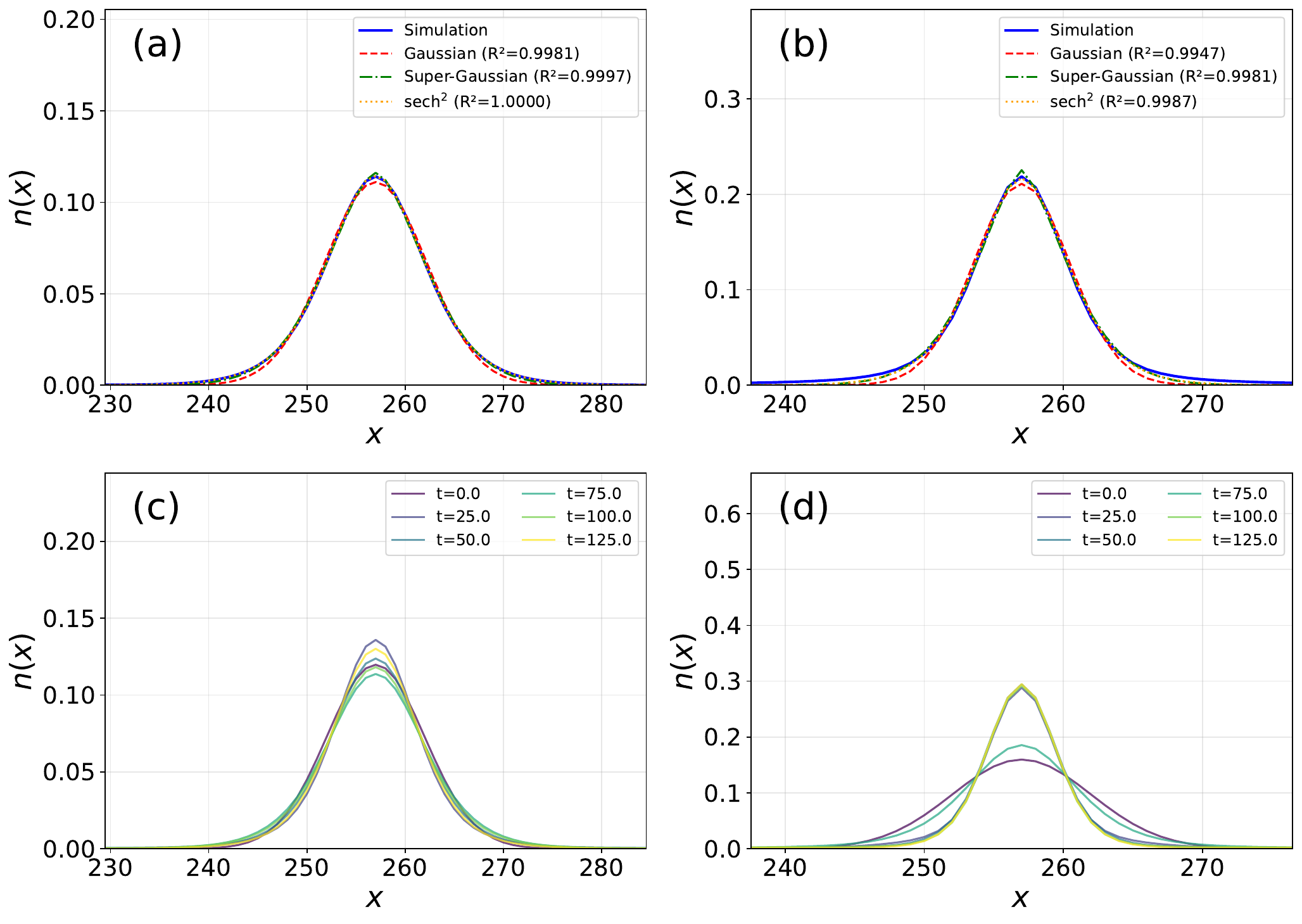}
\caption{Density profile at $t_{\text{eq}}$ for equilibrated (a) homonuclear and (b) heteronuclear droplets (blue solid line) overlaid with their $\text{sech}^2$ (best fit), Gaussian, and super-Gaussian fits. Panels (c) and (d) present the time evolution of the density profiles presented in panels (a) and (b), respectively.}
\label{fig:profile_fits_equilibrated}
\end{figure}

To characterize the equilibrium shape of the droplets, we fit the late-time density profile $n(x, t \to t_{\max})$ of each formed run against three candidate functional forms, all centered at $x_0$ with width parameter $w$ and peak density $n_0$:
\begin{align}
n_{\rm G}(x) &= n_0 \exp\left[-\left(\tfrac{x-x_0}{w}\right)^2\right], \label{eq:gaussian_profile} \\
n_{\rm sech^2}(x) &= n_0 \text{sech}^2\left(\tfrac{x-x_0}{w}\right), \label{eq:sech2_profile}\\
n_{\rm SG}(x) &= n_0 \exp\left[-\left|\tfrac{x-x_0}{w}\right|^p\right], \label{eq:supergaussian_profile}
\end{align}
where the additional free parameter $p$ in super-Gaussian form (\ref{eq:supergaussian_profile}) controls the steepness of the profile boundaries: $p = 2$ recovers the Gaussian, $p < 2$ produces a peaked profile with shallow tails, and $p > 2$ produces a flat-topped profile with sharp boundaries that becomes box-shaped for $p \to \infty$. Each form corresponds to a distinct physical regime: a Gaussian is the natural shape of a weakly-interacting cloud in a harmonic confinement, a $\text{sech}^2$ profile is the bright-soliton solution of the attractive Gross-Pitaevskii equation, and a flat-top super-Gaussian with $p > 2$ is the canonical signature of a quantum droplet in the LHY-saturated regime, where repulsive corrections prevent the central density from increasing further once the droplet is large enough \cite{tylutki2020collective}. Comparing the goodness-of-fit across these three forms therefore probes which regime best describes the droplets in our parameter sweep. We fit each profile by nonlinear least-squares using the Levenberg-Marquardt algorithm and report the coefficient of determination $R^2$ for each form. In Figure \ref{fig:profile_fits_equilibrated}, we present the density profile characterization for the equilibrated profiles best suited by $\text{sech}^2$ for representative homonuclear and heteronuclear cases.

All three models fit extremely well across the dataset: In all 93 formed droplets, the median value $R^2$ of any individual model exceeds 0.997, and in 85 of 93 runs, the three models differ in $R^2$ by less than 0.01. This near-degeneracy partly reflects the smoothness of the simulated density profiles relative to the parametric flexibility of even three-parameter functional forms, and partly reflects the structural relationship between the candidates: the super-Gaussian form in Equation (\ref{eq:supergaussian_profile}) reduces to the Gaussian form in Equation (\ref{eq:gaussian_profile}) at $p = 2$, so super-Gaussian $R^2$ values are guaranteed to satisfy $R^2_{\rm SG} \geq R^2_{\rm G}$
on any data, with strict inequality whenever the data deviate from a perfect Gaussian. Selecting the highest-$R^2$model run-by-run therefore biases the comparison toward super-Gaussian, particularly for runs where the underlying profile is essentially Gaussian and the super-Gaussian fit absorbs the residuals into a small departure of $p$ from 2. We will report both the formal model-selection results and a complementary assessment that controls for this structural bias.

Across all 93 formed droplets, the formal assignment of argmax-$R^2$ yields a nearly even split between super-Gaussian forms (48 cases) and $\text{sech}^2$ (45 cases), with Gaussian never winning by construction. The discrimination is statistically marginal: 80 of the 93 winning assignments are decided by a margin $R^2$ below 0.001 between the best model and the runner-up, which implies that the formal argmax distribution is not evidence that one functional form is preferred over another at the level of the entire dataset. A more selective view emerges when we restrict the analysis to the equilibrated subset (15 droplets), where the late-time density profile reflects a robust stationary state rather than a snapshot of an unfinished relaxation process. Within this subset, the form $\text{sech}^2$ wins in 14 of 15 cases, with a mean fit quality $R^2_{\rm sech^2} = 0.99970 \pm 0.00042$. The super-Gaussian form, although it can in principle reproduce the $\text{sech}^2$ shape with appropriate $p$ and would benefit from its extra degree of freedom, is nevertheless slightly worse than $\text{sech}^2$ in 14 of these 15 runs ($R^2_{\rm SG} = 0.99942 \pm 0.00057$). The single exception is a run for which the super-Gaussian wins by $\Delta R^2 = 6.46 \times 10^{-4}$, comparable to the typical  difference between $\text{sech}^2$ and super-Gaussian forms within the equilibrated subset and well below any threshold for physical significance. We therefore conclude that, for droplets that have actually reached equilibrium within the simulated time horizon, the density profile is marginally best described by $\text{sech}^2$. 

Across the 14 equilibrated droplets that fit the best by $\text{sech}^2$ form, the maximum density and width of the fit are $\langle n_0 \rangle = 0.1899 \pm 0.0232$ and $\langle w \rangle = 5.15 \pm 0.49$ in lattice units, with a mean fit quality $\langle R^2 \rangle = 0.99973 \pm 0.00041$. The 12 homonuclear droplets had $\langle n_0 \rangle_{\text{homo}} = 0.1854 \pm 0.0221$ and $\langle w \rangle_{\text{homo}} = 5.2757 \pm 0.4113$, while the 2 heteronuclear droplets (both at $m_2/m_1 = 2$) reached $\langle n_0 \rangle_{\text{hetero}} = 0.2169 \pm 0.0003$ and $\langle w \rangle_{\text{hetero}} = 4.3781 \pm 0.0055$. The heteronuclear droplets are therefore narrower and slightly denser than their homonuclear counterparts, a difference consistent with the interpretation that a lighter effective reduced mass $m_1 m_2/(m_1+m_2)$ in the asymmetric case produces more compact bound states, although the small heteronuclear sample size precludes a definitive claim. In all 14 cases the equilibrium width is of order $w \sim 5$, well-resolved by the grid spacing, a regime in which the soliton-like sech$^2$ form is the natural 1D-droplet shape \cite{petrov2016ultradilute}.

\begin{figure}
\centering
\includegraphics[width=\columnwidth]{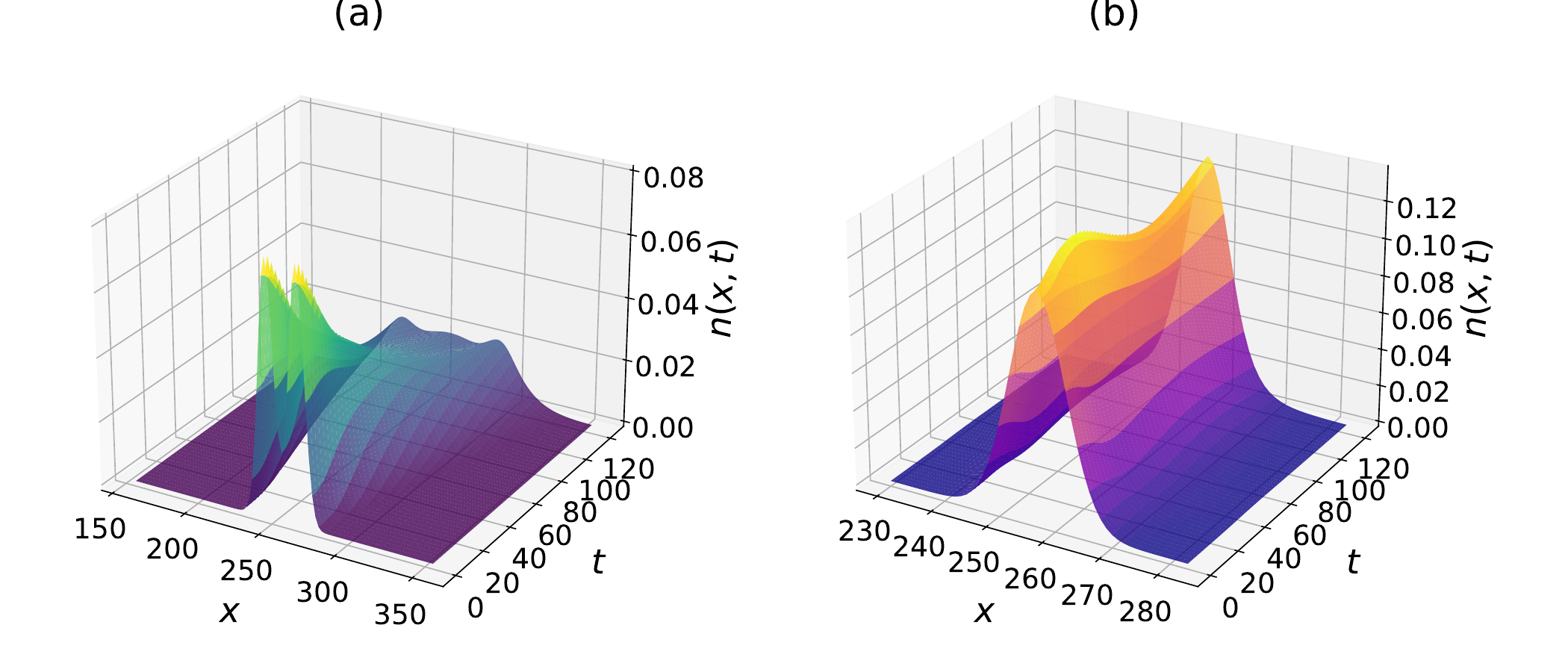}
\caption{Snapshots of the time evolution of (a) an un-equilibrated homonuclear droplet ($\delta U=-0.05$) best fitted by a Super-Gaussian ($R^2=0.98836$) and (b) an equilibrated homonuclear droplet ($\delta U=-0.05$) best fitted by $\text{sech}^2$ ($R^2=0.99998$).}
\label{fig:3d-evolution-homo}
\end{figure}

\begin{figure}
\centering
\includegraphics[width=\columnwidth]{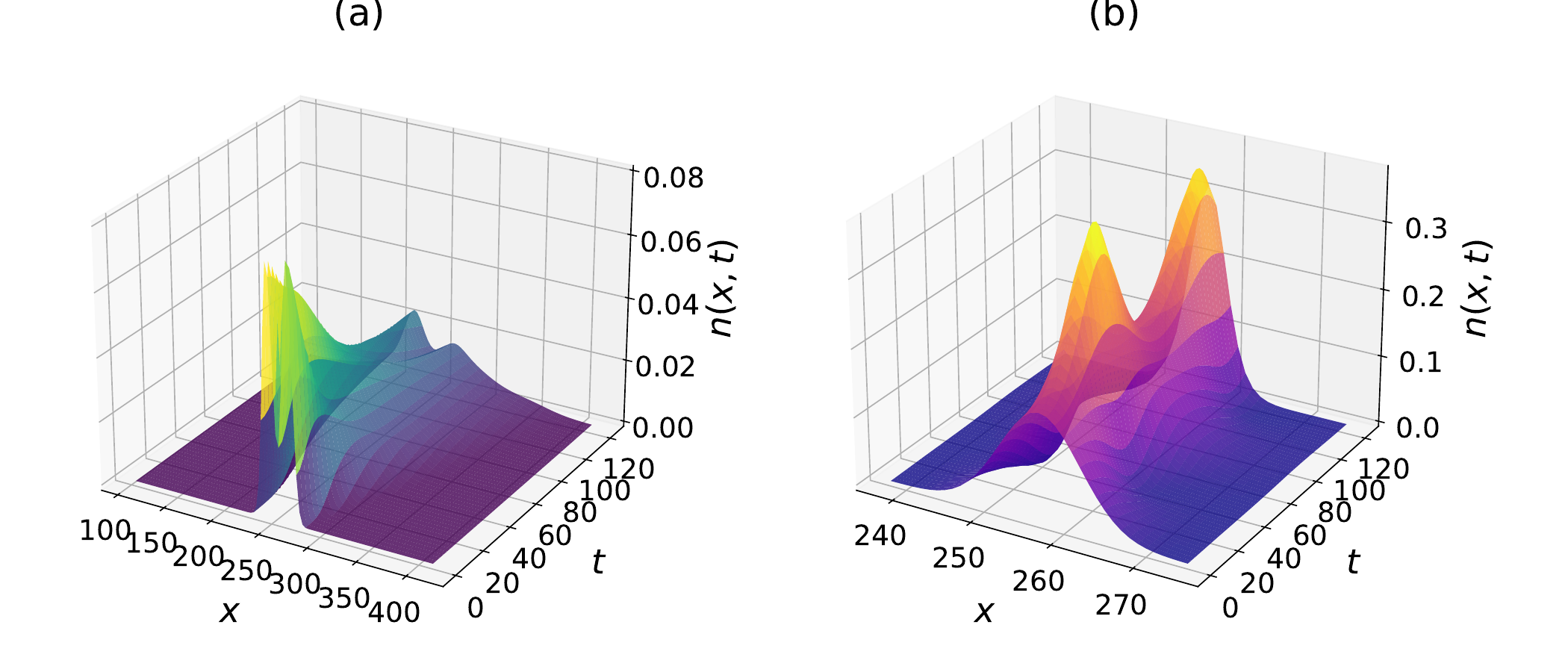}
\caption{Snapshots of the time evolution of (a) an un-equilibrated heteronuclear droplet ($\delta U = -0.05$ and $m_2/m_1=0.5$) best fitted by a Super-Gaussian ($R^2=0.86924$) and (b) an equilibrated heteronuclear droplet ($\delta U=-0.02$ and $m_2/m_1=2$) best fitted by $\text{sech}^2$ ($R^2=0.99874$).}
\label{fig:3d-evolution-hetero}
\end{figure}

This finding contrasts with the formal argmax-$R^2$ result on the full dataset, which is dominated by un-equilibrated runs (78 cases out of 93). Figures \ref{fig:3d-evolution-homo} and \ref{fig:3d-evolution-hetero} present a comparison of time evolution between transient dynamics in un-equilibrated cases best fitted by Super-Gaussian forms and stable equilibrated profiles best fitted by $\text{sech}^2$ forms for homonuclear and heteronuclear droplets, respectively. 

Restricted to the subset that reached droplet formation but not droplet equilibration, the super-Gaussian forms have mean $R^2_{\text{SG}} = 0.98537 \pm 0.04126$ and fit best 47 cases and $\text{sech}^2$ has mean $R^2_{\text{sech}^2}= 0.98168\pm 0.04411$ and fit best 31 cases, which is the inverse of the equilibrated-subset preference. We interpret this as a consequence of incomplete relaxation: profiles sampled at an intermediate point of their dynamical evolution exhibit transient flat-topped or boundary-deformed shapes that the super-Gaussian's adjustable $p$ parameter can absorb but that are not characteristic of the asymptotic droplet state. The dependence of the inferred profile shape on whether the relaxation is complete is itself a manifestation of the equilibration shortfall documented earlier: the limited observation horizon affects not only equilibrium energy and width, but also the inferred shape for the density profile. The preference $\text{sech}^2$ among equilibrated droplets is consistent with the 1D quantum-droplet picture proposed by Petrov and Astrakharchik \cite{petrov2016ultradilute}, in which the soliton-like profile $\text{sech}^2$ emerges from the balance between net mean-field repulsion and attractive LHY corrections. It is also notable that we do not observe the flat-top profile characteristic of larger 1D droplets at high particle numbers, which would correspond to super-Gaussian fits with $p \gg 2$. The flat profile that has been extensively characterized in quantum droplets is controlled by the flatness parameter $\mathcal{F} \propto \delta UN^{2/3}$ \cite{tylutki2020collective} and appears only for large $\gamma>0$. The region of the parameter space we explored in these simulations only covers the range $\vert\mathcal{F}\vert\in[0,1.02115]$, indicating that flat-tops are unlikely to appear in this soliton-like regime.

\section{Conclusions}
\label{conclusions}

We have systematically studied droplet formation using a real-time evolution of quantum droplets in two-component Bose mixtures in one dimension. We integrated extended Gross-Pitaevskii equations, derived from a Bogoliubov formalism including a Lee-Huang-Yang first-order correction in energy. We explicitly integrate their $k$-space integral expressions to avoid errors introduced by analytical approximations. Our aim was to compare homonuclear and heteronuclear systems in a statistical approach by measuring various parameters from its dynamic evolution for different interaction strengths, mass-imbalance and initial conditions. 

Our principal physical conclusion is that droplet formation is robust and mass-symmetric, while droplet equilibration is rare and mass-asymmetric. Droplets form with comparable success rates across homonuclear and heteronuclear configurations, regardless of the mass ratio, but only a small minority reach a stable equilibrium configuration within the simulated timescale. The most striking finding of this work is that this equilibration shortfall is not an arbitrary numerical limitation, but a direct manifestation of weak dissipation in one-dimensional dynamics: the breathing-mode damping timescale substantially exceeds the simulated time horizon. The persistence of breathing oscillations on timescales longer than typical experimental and computational windows is therefore not a peculiarity of our simulations but a general feature of one-dimensional droplet dynamics. We expect this to have direct experimental relevance: preparation protocols that pass through the formation threshold will produce droplets carrying long-lived breathing excitations, and these excitations will dominate the apparent dynamics of the system on experimentally accessible timescales. 

The second main conclusion is methodological. The energetic observables demand a careful choice of reference: on the lattice the total energy is conserved and offset by the free-atom band energy, so that only the binding energy $E_{\text{bind}} = E_{\text{total}} + 2(J_1 N_1 + J_2 N_2)$ carries physical meaning, and its apparent ensemble scatter resolves into deterministic dependence on particle number and mass ratio rather than incomplete relaxation. The characterization of density profiles, in contrast, is genuinely sensitive to relaxation: the full dataset suggests a preference for flat-top profiles, but the equilibrated subset reveals an unambiguous preference for soliton-like shapes $\text{sech}^2$ consistent with the small-droplet limit. We argue that any future statistical study of real-time droplet dynamics should report results separately for the equilibrated and un-equilibrated subsets, as a system stabilized by the LHY mechanism need not have overcome transient formation effects.

Our energy partitioning analysis confirms what has been argued on theoretical grounds but has not been demonstrated in real-time 1D dynamics: the Lee-Huang-Yang contribution is not a perturbative correction to the mean-field binding but {\it is} the binding mechanism itself --- the mean-field interaction energy nearly cancels, and is net repulsive for $\delta U > 0$, while the LHY energy supplies the attraction. The physical distinctness of homonuclear and heteronuclear droplets emerges most strongly here, as the binding deepens monotonically with the mass ratio through the reduced kinetic-energy cost of localization in mass-asymmetric mixtures, and heteronuclear droplets support larger breathing amplitudes through the additional degree of freedom provided by relative motion between components. These features are not present in the homonuclear case and constitute genuine signatures of mass asymmetry in the droplet physics. The self-binding also scales super-extensively with atom number, $|E_{\text{bind}}| \simeq 0.08\, N^{1.6}$, nearly identically at the two interaction strengths sampled. 

The limitations of this analysis suggest natural extensions for future work. The restricted equilibrated subset prevents us from drawing definitive conclusions about the asymptotic density profiles of heteronuclear droplets; a substantial extension of the simulated timescale would address this point. We also relied on the assumption that the system operates near the mean-field cancelation boundary to derive our LHY form, which loses validity for some large $\vert \delta U\vert$ in our dataset. Finally, the parameter regime we explore is the soliton-like small-droplet regime; accessing the flat-top regime characterized by larger particle numbers would connect our dynamical findings to the equilibrium structure of the broader 1D droplet phase diagram.

In concluding this report, we note that the picture that emerges from the results contained herein is that quantum droplets in one dimension are highly stable structures that nevertheless retain coherent collective excitations on timescales characteristic of the surrounding phase space. This unusual combination of structural robustness and dynamical persistence deserves systematic study in both theoretical and experimental contexts. The tools we develop here, in particular, the criteria for droplet formation and equilibration, the use of breathing-mode analysis as a dynamical fingerprint of one-dimensional dissipation, and the use of a variant of the tanh-sinh quadrature for real-time integration should transfer to a wide range of problems in low-dimensional ultracold physics.


\section*{ACKNOWLEDGMENTS}
The authors thank Professors Sandra Denise Prado, Sergio Garcia 
Magalh$\tilde{\rm a}$es and Marcos S\'ergio Figueira da Silva for their insightful comments. Simulations were performed on 4th Generation Intel Xeon Scalable processors using designed x86-64 architecture.

\appendix

\section{Tight-binding model and Crank-Nicolson scheme}
\label{Crank-Nicolson appendix}

In Sect. \ref{section-III} we presented the continuous version of extended Gross-Pitaevskii equations that govern the dynamics of the system studied in this paper. In this Appendix, we describe the choice of discretization used for numerical integration and the algorithm chosen for the task. We use a 1D tight-binding model whose time evolution is described by

\begin{align}
i\hbar &\frac{\psi_i^{n+1} - \psi_i^n}{\Delta t} = -J_1\Bigl(\psi_{i+1}^n + \psi_{i-1}^n\Bigr) \notag \\&+ \Bigl(U_1|\psi_i^n|^2 + U_{12}|\phi_i^n|^2 + \Delta\mu_1^{\text{LHY}}\big|_i^n\Bigr)\psi_i^n, \label{eq:TB1} \\[1ex]
i\hbar &\frac{\phi_i^{n+1} - \phi_i^n}{\Delta t} = -J_2\Bigl(\phi_{i+1}^n + \phi_{i-1}^n\Bigr) \notag \\&+ \Bigl(U_2|\phi_i^n|^2 + U_{12}|\psi_i^n|^2 + \Delta\mu_2^{\text{LHY}}\big|_i^n\Bigr)\phi_i^n, 
\label{eq:TB2}
\end{align}

\noindent
where $\psi_i^n$ ($\phi_i^n$) is the Wannier wavefunction for the first (second) condensate on the $i$-th lattice site at the $n$-th timestep. $J_1$ and $J_2$ denote the hopping terms for the first and second condensates, respectively. $\Delta\mu_1^{\text{LHY}}\big|_i^n$ is the chemical potential contribution due to the LHY correction for the first condensate, evaluated numerically at the site $i$ and at time $n$. Similarly, $\Delta\mu_2^{\text{LHY}}\big|_i^n$ represents the same for the second condensate.

The pure-hopping kinetic term in Eqs.~\eqref{eq:TB1}--\eqref{eq:TB2} differs from the finite-difference form of $-(\hbar^2/2m_\sigma)\,\partial_x^2$ by the constant $-2J_\sigma$ per particle; the two conventions differ only by a global phase and yield identical dynamics, and the constant is accounted for by the energy reference $E_{\text{free}}$ of Eq.~\eqref{eq:ebind}.

Equations~\eqref{eq:TB1} and \eqref{eq:TB2} are integrated using a variant of the Crank-Nicolson scheme \cite{crank1947practical,press2007numerical}. The standard Crank-Nicolson method for the Schr\"odinger equation \cite{weizhu2003numerical,antoine2013computational} averages the right-hand side at times $t_n$ and $t_{n+1}$:
\begin{equation}
i\hbar \frac{\psi_i^{n+1} - \psi_i^n}{\Delta t} = \frac{1}{2}\sum_{k=1}^{N_{\text{sit}}} \bigl(H_{ik}^{(1),n+1}\psi_k^{n+1} + H_{ik}^{(1),n}\psi_k^n\bigr), \label{eq:CN_full}
\end{equation}
for the first condensate and, equivalently, for the second condensate. Here, the effective single-particle Hamiltonian for the first condensate is
\begin{equation}
H_{ik}^{(1),n} = -J_1(\delta_{i,k+1} + \delta_{i,k-1}) + V_{\text{eff},i}^{(1),n}\,\delta_{ik},
\end{equation}
with the effective on-site potential
\begin{equation}
V_{\text{eff},i}^{(1),n} = U_1|\psi_i^n|^2 + U_{12}|\phi_i^n|^2 + \Delta\mu_1^{\text{LHY}}\big|_i^n.
\end{equation}

The effective potential $V_{\text{eff},i}^{(1)}$ (and therefore the effective Hamiltonian $H^{(1)}$)  depends on time through the local densities $n_{1,i}(t) = |\psi_i(t)|^2$ and $n_{2,i}(t) = |\phi_i(t)|^2$. 

A fully implicit Crank-Nicolson scheme as in Eq.~\eqref{eq:CN_full} would require evaluating $H^{n+1} = H(|\psi^{n+1}|^2, |\phi^{n+1}|^2)$, but $\psi^{n+1}$ is the unknown for which we are solving. This creates a nonlinear system that must be solved iteratively at each timestep, significantly increasing the computational cost. To avoid this overhead, we adopt a semi-implicit approach, also known as lagged nonlinearity \cite{sanz2018numerical}: the effective potential is evaluated entirely at the current time $t_n$, yielding the scheme
\begin{equation}
i\hbar \frac{\psi_i^{n+1} - \psi_i^n}{\Delta t} = \frac{1}{2}\sum_{k} H_{ik}^{n}\bigl(\psi_k^{n+1} + \psi_k^n\bigr), \label{eq:CN_semi}
\end{equation}
where $H^n \equiv H(|\psi^{n}|^2, |\phi^{n}|^2)$. This can be rewritten in matrix form as
\begin{equation}
\Bigl[I + \frac{i\Delta t}{2\hbar}H^n\Bigr]\psi^{n+1} = \Bigl[I - \frac{i\Delta t}{2\hbar}H^n\Bigr]\psi^n. \label{eq:CN_matrix}
\end{equation}
Since $H^n$ is known from the previous timestep, this is a linear tridiagonal system, which we solve by the Thomas algorithm \cite{thomas1949elliptic,golub2013matrix}. The same treatment is applied to the second condensate.

We now derive explicitly the truncation error introduced by the semi-implicit approximation. The right-hand side of Eq. \eqref{eq:CN_semi} is
\begin{equation}
\frac{1}{2}H^n\bigl(\psi^{n+1} + \psi^n\bigr) = H^n \cdot \frac{\psi^{n+1} + \psi^n}{2}.
\label{right-hand-side}
\end{equation}

Let $t_{n+1/2} = t_n + \Delta t/2$ denote the temporal midpoint and let $\psi_i^* \equiv \psi_i(t_{n+1/2})$ and $H^* \equiv H(t_{n+1/2})$ be the exact wave function and Hamiltonian at this point. Expanding around the midpoint, we have the following:
\begin{align}
\psi_i^n &= \psi_i^* - \frac{\Delta t}{2}\dot{\psi}_i^* + \frac{(\Delta t)^2}{8}\ddot{\psi}_i^* + \mathcal{O}((\Delta t)^3), \label{eq:psi_n}\\[1ex]
\psi_i^{n+1} &= \psi_i^* + \frac{\Delta t}{2}\dot{\psi}_i^* + \frac{(\Delta t)^2}{8}\ddot{\psi}_i^* + \mathcal{O}((\Delta t)^3), \label{eq:psi_n1}\\[1ex]
H^n &= H^* - \frac{\Delta t}{2}\dot{H}^* + \mathcal{O}((\Delta t)^2), \label{eq:H_n}\\[1ex]
H^{n+1} &= H^* + \frac{\Delta t}{2}\dot{H}^* + \mathcal{O}((\Delta t)^2). \label{eq:H_n1}
\end{align}
From Eqs. \eqref{eq:psi_n}-\eqref{eq:psi_n1}, the average wavefunction is
\begin{equation}
\frac{\psi^{n+1} + \psi^n}{2} = \psi^* + \frac{(\Delta t)^2}{8}\ddot{\psi}^* + \mathcal{O}((\Delta t)^3).
\label{eq:psi_avg}
\end{equation}
This remains second-order accurate. Subtracting Eq. \eqref{eq:psi_n} from Eq. \eqref{eq:psi_n1}, we find:
\begin{equation}
\frac{\psi_i^{n+1} - \psi_i^n}{\Delta t} = \dot{\psi}_i^* + \mathcal{O}((\Delta t)^2). 
\label{eq:LHS}
\end{equation}

The Hamiltonian can be decomposed into linear and nonlinear parts as $H = H_{\text{linear}} + H_{\text{nonlinear}}$, where $H_{\text{linear}}$ contains the hopping terms which are independent of densities and $H_{\text{nonlinear}}$ contains the mean-field and LHY terms. The linear part satisfies $H_{\text{linear}}^n = H_{\text{linear}}^* = H_{\text{linear}}^{n+1}$ exactly. For the nonlinear part, from Eq. \eqref{eq:H_n}:
\begin{equation}
H_{\text{nonlinear}}^n = H_{\text{nonlinear}}^* - \frac{\Delta t}{2}\dot{H}_{\text{nonlinear}}^* + \mathcal{O}((\Delta t)^2). \label{eq:Hnonlinear_expansion}
\end{equation}
Inserting this into Eq. \eqref{right-hand-side}, we reach:
\begin{align}
&H^n \cdot \frac{\psi^{n+1} + \psi^n}{2} \nonumber\\&= \bigl(H_{\text{linear}} + H_{\text{nonlinear}}^*  - \frac{\Delta t}{2}\dot{H}_{\text{nonlinear}}^*\bigr)\bigl(\psi^* + \mathcal{O}((\Delta t)^2)\bigr) \nonumber\\ &+ \mathcal{O}((\Delta t)^2) 
= H^*\psi^* - \frac{\Delta t}{2}\dot{H}_{\text{nonlinear}}^*\psi^* + \mathcal{O}((\Delta t)^2). \label{eq:RHS_semi}
\end{align}
In particular, the LHY correction $\Delta\mu_1^{\text{LHY}}$ to the chemical potential of the first condensate is a smooth functional of densities $n_1$ and $n_2$:
\begin{equation}
\Delta\mu_1^{\text{LHY}}\big|_i^n = \mathcal{F}\bigl(n_{1,i}^n, n_{2,i}^n;\, U_{1}, U_{2}, U_{12}, m_1, m_2\bigr),
\end{equation}
where $\mathcal{F}$ denotes the integral over quasi-momentum computed via exp-sinh quadrature (see Appendix \ref{tanh-sinh-quadrature}). The exp-sinh quadrature evaluates this integral to high precision for given density values; it introduces no time-stepping error since it computes an essentially exact function of its inputs at each instant. Since $\mathcal{F}$ is smooth, we can expand:
\begin{align}
\Delta\mu_1^{\text{LHY}}\big|_i^n &= \Delta\mu_1^{\text{LHY}}\big|_i^* - \frac{\Delta t}{2}\left(\frac{\partial \mathcal{F}}{\partial n_1}\dot{n}_{1,i}^* + \frac{\partial \mathcal{F}}{\partial n_2}\dot{n}_{2,i}^*\right) \nonumber \\
&\quad + \mathcal{O}((\Delta t)^2).
\end{align}

The exact solution satisfies $i\hbar\dot{\psi}^* = H^*\psi^*$. Comparing the left-hand side of the numerical scheme in Eq. \eqref{eq:CN_semi}, which is presented in Eq. \eqref{eq:LHS}, with Eq. \eqref{eq:RHS_semi}, which represents the right-hand side of Eq. \eqref{eq:CN_semi}, we find:
\begin{align}
i\hbar\Bigl(\dot{\psi}^* + \mathcal{O}((\Delta t)^2)\Bigr) &= H^*\psi^* - \frac{\Delta t}{2}\dot{H}_{\text{nonlinear}}^*\psi^* + \mathcal{O}((\Delta t)^2).
\end{align}
Using $i\hbar\dot{\psi}^* = H^*\psi^*$, the local truncation error is
\begin{equation}
\tau_{\text{local}} = \frac{\Delta t}{2i\hbar}\dot{H}_{\text{nonlinear}}^*\psi^* + \mathcal{O}((\Delta t)^2) = \mathcal{O}(\Delta t). \label{eq:local_error}
\end{equation}

The error term $\dot{H}_{\text{nonlinear}}^*\psi^*$ is proportional to the rate of change of the effective potential:
\begin{equation}
\dot{H}_{\text{nonlinear}}^* = U_1 \frac{d|\psi^*|^2}{dt} + U_{12}\frac{d|\phi^*|^2}{dt} + \frac{d\Delta\mu_1^{\text{LHY}}}{dt},
\end{equation}
where $\frac{d|\psi|^2}{dt} = 2\,\text{Re}\bigl(\psi^*\dot{\psi}^\dagger\bigr)$ represents the local probability current. For slowly varying density profiles, such as quasi-equilibrium dynamics, $\dot{H}_{\text{nonlinear}}^*$ is small, and the error is suppressed even though it is formally $\mathcal{O}(\Delta t)$.

For a scheme with local truncation error $\mathcal{O}((\Delta t)^p)$, the global error is typically $\mathcal{O}((\Delta t)^{p})$. Thus, the scheme is first-order accurate in time, with the dominant error arising from the nonlinear sector. Despite the reduced temporal accuracy, the semi-implicit scheme retains crucial stability properties \cite{caliari2009high}. The update can be written as
\begin{equation}
\psi^{n+1} = \Bigl[I + \frac{i\Delta t}{2\hbar}H^n\Bigr]^{-1}\Bigl[I - \frac{i\Delta t}{2\hbar}H^n\Bigr]\psi^n \equiv U^n \psi^n.
\end{equation}
When $H^n$ is Hermitian, as is the case for real interaction strengths, the propagator $U^n$ is unitary. Since $(I \pm \frac{i\Delta t}{2\hbar}H^n)^\dagger = (I \mp \frac{i\Delta t}{2\hbar}H^n)$ for the Hermitian $H^n$, we have
\begin{equation}
(U^n)^\dagger = \left[I + \frac{i\Delta t}{2\hbar}H^n\right]\left[I - \frac{i\Delta t}{2\hbar}H^n\right]^{-1}.
\end{equation}
Therefore, one can easily demonstrate that $(U^n)^\dagger U^n = I$, because all four matrices composing the formulas for $U^n$ and $(U^n)^\dagger$ commute (as polynomials in $H^n$) and thus can be rearranged to cancel pairwise. This ensures exact norm conservation $\|\psi^{n+1}\|^2 = \|\psi^n\|^2$ at every time step, regardless of $\Delta t$. The scheme is therefore unconditionally stable \cite{bao2004computing}, as no restriction on the timestep is required for numerical stability. 

In our simulations, we use $\Delta t = 0.001$ in dimensionless units. Even with formally first-order accuracy in nonlinear terms, the error prefactor $\frac{1}{2}\dot{H}_{\text{nonlinear}}^*\psi^*$ remains small for the parameter regimes studied, where density profiles evolve smoothly. Norm conservation, guaranteed exactly by the unitarity of the propagator, is verified at every time step. The total energy, which the semi-implicit scheme conserves only up to the $O(\Delta t)$ truncation error derived above, drifts by less than $5.2\times 10^{-2}$ over the full simulation time in every run of the campaign, with a median drift of $1.4\times 10^{-3}$, confirming that the lagged-nonlinearity error remains small in practice.

Integration of $\Delta\mu_1^{\text{LHY}}\big|_i^n$ and $\Delta\mu_2^{\text{LHY}}\big|_i^n$ is carried out numerically for every lattice site and every timestep using the exp-sinh quadrature, described in Appendix \ref{tanh-sinh-quadrature}.

\section{Tanh-sinh quadrature and exp-sinh quadrature}
\label{tanh-sinh-quadrature}
 Due to the consideration of heteronuclearity in our system, the Lee-Huang-Yang correction to the chemical potential, in Eq. (\ref{chemical-potential-integral-in-k}), requires numerical evaluation of an integral over quasi-momentum arising from the Bogoliubov spectrum, as no closed-form expression for it exists. In this Appendix, we describe the double-exponential method used for this purpose and justify its suitability for our problem.

The LHY corrections at each lattice site $i$ involve integrals of the form
\begin{equation}
\Delta\mu^{\text{LHY}}_\sigma\big|_i = \int_0^{\infty} f_\sigma(k;\, n_{1,i}, n_{2,i})\, dk,
\label{eq:LHY_integral}
\end{equation}
where $\sigma \in \{1,2\}$ labels the condensate component, $k$ is the quasi-momentum, and the integrand $f_\sigma$, representing the derivative of the Bogoliubov zero-point sum including its counterterm, Eq.~(\ref{chemical-potential-integral-in-k}), depends on the local densities $n_{1,i} = |\psi_i|^2$ and $n_{2,i} = |\phi_i|^2$, which are time-dependent. The same construction applies to the energy density itself, Eq. (\ref{energy-density-integral}), whose numerical evaluation is discussed at the end of this Appendix. The integrand is smooth on $(0,\infty)$ apart from the phonon regime near $k = 0$, where it behaves as $f(k) \sim |k|^p$ for some $p > 0$, and it decays only algebraically, as $1/k^2$, at large $k$ once the counterterm cancellation has taken place; in the mean-field-unstable window of quasi-momenta, where $\omega_-^2 < 0$, the modes are excluded from the integration, implementing the neglect of dynamical instabilities stated in Section~\ref{section-III}.

Standard quadrature methods such as Gauss-Legendre can struggle with endpoint stiffness \cite{davis2007methods} and require many nodes to achieve high accuracy. The tanh-sinh quadrature is specifically designed to handle such integrands efficiently. It was introduced by Takahashi and Mori in 1974 \cite{tanh-sinh-quadrature}, and it belongs to the family of double-exponential methods \cite{double-exponential-method}. The core idea is to transform the integration variable so that the integrand, including any endpoint singularities or stiffness, is mapped to a new function that decays double-exponentially fast, allowing efficient truncation of the integration domain.

We begin by considering a generic integral over the finite region $[-1, 1]$:
\begin{equation}\label{genericintegral}
I = \int_{-1}^{1} g(x) \, dx.
\end{equation}
The tanh-sinh transformation introduces a new variable $t \in (-\infty, \infty)$ through
\begin{equation}
x = \varphi(t) \equiv \tanh\!\bigl(c \cdot \sinh(t)\bigr), \label{eq:tanh_sinh_transform}
\end{equation}
where $c$ is a constant, conventionally taken to be $c = \pi/2$. The integral presented in \eqref{genericintegral} becomes
\begin{equation}
I = \int_{-\infty}^{\infty} g\bigl(\varphi(t)\bigr) \, \varphi'(t) \, dt.
\label{eq:transformed_integral}
\end{equation}
The key property of this transformation is the asymptotic behavior of $\varphi'(t)$, which scales as $\sim \exp\!\bigl(-c \cdot e^{|t|}\bigr)$ as $|t| \to \infty.$ This double-exponential decay ensures that the transformed integrand $g(\varphi(t))\,\varphi'(t)$ vanishes extremely rapidly as $|t| \to \infty$, even if $g(x)$ has integrable singularities or large derivatives at $x = \pm 1$. Consequently, the infinite integral in Eq. \eqref{eq:transformed_integral} can be truncated to a finite interval $[-T, T]$ with negligible error for modest values of $T$.

A complementary perspective is that the transformation concentrates the quadrature nodes near the end points $x = \pm 1$. As $t \to +\infty$, we have $\varphi(t) \to 1^-$, and the nodes $x_k = \varphi(t_k)$ cluster densely near $x = 1$. Similarly for $x = -1$ as for $t \to -\infty$. This dense sampling near endpoints is precisely what is needed to resolve endpoint stiffness or weak singularities.

The integral in Eq.~\eqref{eq:LHY_integral}, however, runs over the semi-infinite domain $(0,\infty)$, for which the appropriate member of the same double-exponential family is the exp-sinh transformation,
\begin{equation}
k = \phi(t) \equiv \exp\!\Bigl(\frac{\pi}{2}\sinh(t)\Bigr),
\qquad
dk = \frac{\pi}{2}\cosh(t)\,\phi(t)\, dt,
\label{eq:exp_sinh_transform}
\end{equation}
with $t \in (-\infty,\infty)$. It plays for $(0,\infty)$ the role that Eq.~\eqref{eq:tanh_sinh_transform} plays for $[-1,1]$: as $t \to -\infty$ the nodes $k_j = \phi(t_j)$ accumulate double-exponentially toward $k = 0^+$, resolving the phonon regime with no special treatment, while as $t \to +\infty$ they spread double-exponentially toward large $k$, covering the slowly decaying $1/k^2$ tail with few nodes. In the transformed variable the integrand decays double-exponentially in $|t|$, so the trapezoidal rule applies with the same spectral accuracy as in the finite-interval case: for integrands that decay double-exponentially, the trapezoidal rule achieves spectral convergence \cite{bailey2005comparison}, faster than any polynomial. This remarkable property arises because the Euler-Maclaurin error terms, which normally limit the accuracy of trapezoidal sums, are exponentially small \cite{trefethen2014exponentially} when the integrand and all its derivatives vanish rapidly at the ends of the integration range.

The transformed integral is discretized on the uniform grid $t_j = j h$, $j \in \mathbb{Z}$, with full trapezoidal weights,
\begin{equation}
\int_0^{\infty} f(k)\, dk \;\approx\; h \sum_{|t_j| \le T} w_j\, f(k_j),
\qquad
w_j = \frac{\pi}{2}\cosh(t_j)\, k_j,
\end{equation}
truncated to the fixed window $T = 3$, for which $k$ spans $[1.5\times10^{-7},\, 6.8\times10^{6}]$ and the window-truncation error is below $10^{-6}$ of the integral. The step is refined by successive halving, $h_\ell = h_0/2^{\ell}$ with $h_0 = 0.44$ and $\ell = 0, 1, \ldots, 14$, and convergence is monitored by comparing consecutive refinement levels, jointly for the three integrals evaluated at each site (the two chemical-potential corrections and the energy density), through both an absolute and a relative criterion. With the conservative tolerances employed ($10^{-20}$, below double precision), the refinement proceeds to the finest level in practice, $h_{14} \approx 2.7\times10^{-5}$, corresponding to $\approx 2.2\times10^{5}$ nodes.

For smooth integrands with at most integrable endpoint singularities, the double-exponential method achieves an error that scales as
\begin{equation*}
\text{error} \sim \exp\!\Bigl(-\frac{c' N}{\log N}\Bigr)
\end{equation*}
for some constant $c' > 0$, where $N$ is the number of nodes.

The same quadrature evaluates the energy density of Eq. (\ref{energy-density-integral}); its integrand, however, requires one algebraic precaution for the sake of numerical stability. Differentiation with respect to $n_\sigma$ removes the $O(k^4)$ terms of the Bogoliubov branches analytically, so the chemical-potential integrand of Eq. (\ref{chemical-potential-integral-in-k}) subtracts only quantities of order $U_\sigma n_\sigma$ at large $k$ and is numerically benign; the energy-density integrand, $f_E = \omega_+ + \omega_- - \hbar^2k^2/2m_1 - \hbar^2k^2/2m_2 - U_1 n_1 - U_2 n_2$, performs the corresponding cancellation numerically, subtracting quantities of order $k^2/2 \sim 10^{13}$ near the upper end of the integration window to leave a remainder of order $10^{-15}$, a loss of some 28 significant digits, beyond the reach of double precision \cite{goldberg1991what}. We therefore rearrange the integrand so that the cancellation is carried out exactly, in the algebra. Writing $f_E = (S^2 - K^2)/(S + K)$, with $S = \omega_+ + \omega_-$ and $K = e_A + e_B + a' + b'$, where $e_\sigma = \hbar^2 k^2/2m_\sigma$, $a' = U_1 n_1$, $b' = U_2 n_2$, and $c^2 = U_{12}^2\, n_1 n_2$, the sum and product of the roots of the Bogoliubov quadratic give $S^2 = e_A^2 + e_B^2 + 2e_A a' + 2e_B b' + 2\,e_A e_B \sqrt{1+X}$, with
\begin{equation}
X = \frac{2a'}{e_A} + \frac{2b'}{e_B} + \frac{4\,(a'b' - c^2)}{e_A e_B}.
\end{equation}
Rationalizing $\sqrt{1+X} - 1 = X/Y$, with $Y = 1 + \sqrt{1+X}$, all terms of order $k^4$ and $k^2$ then cancel identically, leaving
\begin{align}
&f_E = \frac{1}{\,\omega_+ + \omega_- + e_A + e_B + a' + b'\,}\\\times
&\left[\,-2\,(e_A b' + e_B a')\,\frac{X}{Y^{2}} \;+\; \frac{8\,(a'b'-c^{2})}{Y} \;-\; (a'+b')^{2}\right],
\label{eq:stable_fE}
\end{align}
in which every subtraction involves quantities of order $U n$ or smaller, so that the rounding error per node is again of order machine epsilon. The identity $1 + X = \omega_+^{2}\omega_-^{2}/(e_A e_B)^{2}$ shows, in addition, that the condition $1 + X \le 0$ reproduces exactly the exclusion of the mean-field-unstable window. Eq. (\ref{energy-density-integral}) and Eq.~\eqref{eq:stable_fE} are the same function of $k$ at every quasi-momentum, with only their floating-point conditioning differing, such that the stable form agrees with the closed-form result of the homonuclear limit to better than $10^{-6}$ on the grid described above. All energy densities and total energies reported in this work are computed with Eq.~\eqref{eq:stable_fE}.


\bibliographystyle{apsrev4-1}
\bibliography{references}

\end{document}